\documentclass[aps,prx,showpacs,amssymb,amsart,superscriptaddress,nofootinbib,twocolumn,longbibliography]{revtex4-1}

\usepackage{amsmath,amsfonts,amsthm,amssymb}
\usepackage{braket}
\usepackage{float}
\usepackage{setspace}
\usepackage{amsmath}
\usepackage{appendix}
\usepackage[ansinew]{inputenc}
\usepackage{bbm}
\usepackage{bm}
\usepackage{amsbsy}
\usepackage{dsfont} 
\usepackage{graphicx} 
\usepackage{epsfig}
\usepackage{epstopdf}
\usepackage{dsfont}
\usepackage{color}
\usepackage[bookmarks=true, colorlinks,breaklinks]{hyperref}
\makeatletter
\newcommand\org@hypertarget{}
\let\org@hypertarget\hypertarget
\renewcommand\hypertarget[2]{%
  \Hy@raisedlink{\org@hypertarget{#1}{}}#2%
  }
\makeatother

\usepackage[figure,table]{hypcap}
\usepackage{enumerate}
\usepackage[resetlabels]{multibib}

\hypersetup{
	bookmarksnumbered,
	pdfstartview={FitH},
	citecolor={darkgreen},
	linkcolor={darkred},
	urlcolor={darkblue},
	pdfpagemode={UseOutlines}}
\definecolor{darkgreen}{RGB}{50,190,50}
\definecolor{darkblue}{RGB}{0,0,190}
\definecolor{darkred}{RGB}{238,0,0}
\usepackage{soul}
\newcommand{\id}{\mathbbm{1}}

\newcommand{\expval}[1]{\ensuremath{\left\langle\right.\! #1 \!\left.\right\rangle}}

\newcommand{\pr}{^{\prime}}

\newcommand{\subtiny}[3]{\ensuremath{_{\hspace{#1 pt}\protect\raisebox{#2 pt}{\tiny{$ #3$}}}}}
\newcommand{\suptiny}[3]{\ensuremath{^{\hspace{#1 pt}\protect\raisebox{#2 pt}{\tiny{$ #3$}}}}}

\DeclareMathOperator{\diag}{diag}

\newcommand{\tr}{\textnormal{Tr}}
\newcommand{\djj}{d\kern-0.4em\char"16\kern-0.1em}
\usepackage{filecontents}

\begin{document}

%
\title{Observation of Entangled States of a Fully Controlled 20-Qubit System}
\author{Nicolai Friis}
\thanks{These authors contributed equally to this work.}
\affiliation{Institute for Quantum Optics and Quantum Information - IQOQI Vienna, Austrian Academy of Sciences, Boltzmanngasse 3, 1090 Vienna, Austria}
\author{Oliver Marty}
\thanks{These authors contributed equally to this work.}
\affiliation{Institut f\"ur Theoretische Physik and IQST, Albert-Einstein-Allee 11, Universit\"at Ulm, 89069 Ulm, Germany}
\author{Christine Maier}
\affiliation{Institute for Quantum Optics and Quantum Information, Austrian Academy of Sciences, Technikerstra{\ss}e 21a, 6020 Innsbruck, Austria}
\affiliation{Institut f\"ur Experimentalphysik, Universit{\"a}t Innsbruck, Technikerstra{\ss}e 25, 6020 Innsbruck, Austria}
\author{Cornelius Hempel}
\thanks{{\it Current address}: ARC Centre for Engineered Quantum Systems, School of Physics, The University of Sydney, 2006 NSW, Australia}
\affiliation{Institute for Quantum Optics and Quantum Information, Austrian Academy of Sciences, Technikerstra{\ss}e 21a, 6020 Innsbruck, Austria}
\affiliation{Institut f\"ur Experimentalphysik, Universit{\"a}t Innsbruck, Technikerstra{\ss}e 25, 6020 Innsbruck, Austria}
\author{Milan Holz{\"a}pfel}
\affiliation{Institut f\"ur Theoretische Physik and IQST, Albert-Einstein-Allee 11, Universit\"at Ulm, 89069 Ulm, Germany}
\author{Petar Jurcevic}
\affiliation{Institute for Quantum Optics and Quantum Information, Austrian Academy of Sciences, Technikerstra{\ss}e 21a, 6020 Innsbruck, Austria}
\affiliation{Institut f\"ur Experimentalphysik, Universit{\"a}t Innsbruck, Technikerstra{\ss}e 25, 6020 Innsbruck, Austria}
\author{Martin B. Plenio}
\affiliation{Institut f\"ur Theoretische Physik and IQST, Albert-Einstein-Allee 11, Universit\"at Ulm, 89069 Ulm, Germany}
\author{Marcus Huber}
\affiliation{Institute for Quantum Optics and Quantum Information - IQOQI Vienna, Austrian Academy of Sciences, Boltzmanngasse 3, 1090 Vienna, Austria}
\author{Christian Roos}
\affiliation{Institute for Quantum Optics and Quantum Information, Austrian Academy of Sciences, Technikerstra{\ss}e 21a, 6020 Innsbruck, Austria}
\author{Rainer Blatt}
\affiliation{Institute for Quantum Optics and Quantum Information, Austrian Academy of Sciences, Technikerstra{\ss}e 21a, 6020 Innsbruck, Austria}
\affiliation{Institut f\"ur Experimentalphysik, Universit{\"a}t Innsbruck, Technikerstra{\ss}e 25, 6020 Innsbruck, Austria}
\author{Ben Lanyon}
\email{ben.lanyon@uibk.ac.at}
\affiliation{Institute for Quantum Optics and Quantum Information, Austrian Academy of Sciences, Technikerstra{\ss}e 21a, 6020 Innsbruck, Austria}

\begin{abstract}
We generate and characterise entangled states of a register of 20 individually controlled qubits, where each qubit is encoded into the electronic state of a trapped atomic ion. Entanglement is generated amongst the qubits during the out-of-equilibrium dynamics of an Ising-type Hamiltonian, engineered via laser fields. Since the qubit-qubit interactions decay with distance, entanglement is generated at early times predominantly between neighbouring groups of qubits. We characterise entanglement between these groups by designing and applying witnesses for genuine multipartite entanglement. Our results show that, during the dynamical evolution, all neighbouring qubit pairs, triplets, most quadruplets, and some quintuplets simultaneously develop genuine multipartite entanglement. Witnessing genuine multipartite entanglement in larger groups of qubits in our system remains an open challenge.
\end{abstract}
\maketitle

\section{Introduction}\label{sec:introduction}
\vspace*{-2mm}

The ability to generate quantum entanglement~\cite{PlenioVirmani2007, HorodeckiEntanglementReview2009} between large numbers of spatially-separated and individually-controllable quantum systems \textemdash\ such as qubits \textemdash\ is of fundamental importance to a broad range of current research endeavours, including studies of nonlocality~\cite{BrunnerCavalcantiPironioScaraniWehner2014}, quantum computing~\cite{NielsenChuang2000}, quantum simulation~\cite{Lloyd1996}, quantum communication~\cite{GisinRibordyTittelZbinden2002, Kimble08}, and quantum metrology~\cite{WinelandBollingerItanoMooreHeinzen1992, HuelgaMacchiavelloPellizzariEkertPlenioCirac1997, GiovannettiLloydMaccone2006, FriisOrsucciSkotiniotisSekatskiDunjkoBriegelDuer2017}. For example, in order for quantum computers and simulators to go beyond the capabilities of conventional computers, large amounts of entanglement (or other quantum correlations) must be generated between their components~\cite{Vidal2003}.

As such, there is an ongoing effort to generate and characterise entangled states of increasing numbers of qubits, in systems which permit preparation of arbitrary initial states, the control of interactions between constituent particles, and readout of individual sites. In such systems, the largest number of qubits entangled to date is 14, achieved in a trapped-ion system~\cite{MonzEtAl2011}, followed by 10 entangled superconducting qubits~\cite{SongEtAl2017}, and 10 entangled photonic qubits~\cite{WangEtAl2016}.

Since every qubit added to an experimental system doubles the Hilbert space dimension in which the collective quantum state is described, the task of characterisation of an unknown state in the laboratory can soon become a significant challenge. Indeed, all generated entangled states of more than 6 qubits to date have been of a highly symmetric form, such as Greenberger-Horne-Zeilinger (GHZ) or W states, for which efficient characterisation techniques exist~\cite{GuehneToth2009}. How to generate and detect more complex multiqubit entangled states remains an open challenge.

In this paper, we report on the deterministic generation of complex entangled states of 20 trapped-ion qubits and their partial characterisation via custom-built witnesses for genuine multipartite entanglement (GME). Our states are complex in the sense that they are generated during quench dynamics of an engineered many-body Hamiltonian and their exact description requires specifying a number of parameters that grows exponentially in the number of qubits involved. Each qubit in our system can be, and is in this work, individually manipulated and read out, as required for universal quantum computation and quantum simulation.

This paper is structured as follows. Section~\ref{sec:experimental setup} presents the experimental system and explains how the $20$-qubit quantum states are generated and measured in the laboratory. Section~\ref{sec:initial results} presents results of basic properties measured for the generated $20$-qubit states, using established methods. Section~\ref{sec:Vienna witnesses main text} introduces and applies analytically derived GME witnesses to reveal genuine tripartite entanglement in all groups of $3$ neighbouring qubits. To go beyond tripartite correlations, we then turn to more computationally demanding witnesses in Sec.~\ref{sec:Ulm witnesses main text}, which enable GME to be detected in groups of up to $5$ neighbouring qubits. Finally, we discuss our results and possible future directions in Sec.~\ref{sec:discussion}.

\vspace*{-2mm}
\section{Experimental Setup}\label{sec:experimental setup}
\vspace*{-2mm}

Our register of $N=20$ qubits is realized using a 1D string of $^{40}$Ca$^+$ ions confined in a linear Paul trap, with axial (radial) centre-of-mass vibrational frequency of 220~kHz (2.712~MHz)~\cite{HempelPHD}. A qubit is encoded into two long-lived states of the outer valence electron in each ion. That is, the computational basis states of the qubits are chosen as $\ket{0}=\ket{S_{J{=}1/2,m_j{=}1/2}}$,  $\ket{1}=\ket{D_{J{=}5/2,m_j{=}5/2}}$, which are connected by an electric quadrupole transition at 729~nm~\cite{SchindlerEtAl2013}.

Under the influence of laser-induced forces that off-resonantly drive all 40 transverse normal vibrational modes of the ion string, the interactions between the qubits are well described by an ``$XY$" model in a dominant transverse field~\cite{LanyonEtAl2017, JurcevicPHD, KimChangIslamKorenblitDuanMonroe2009}, with Hamiltonian
\begin{align}
    H_{XY} & =\hbar \sum_{i<j}J_{ij}(\sigma_i^+\sigma_j^-{+}\sigma_i^-\sigma_j^+)+\hbar B\sum_{j}\sigma_j^z.
\end{align}
Here $J_{ij}$ is an $N \times N$ qubit-qubit coupling matrix, $\sigma_i^{+}$ ($\sigma_i^{-}$) is the qubit raising (lowering) operator for qubit $i$, $B$ is the transverse field strength ($B\gg \max\{|J_{ij}|\}$) and $\sigma_j^z\equiv Z_{j}$ is the Pauli $Z$ matrix for qubit $j$ with eigenvectors satisfying  $Z\ket{0}=-\ket{0}$ and $Z\ket{1}=\ket{1}$. Interactions reduce approximately with a power law $J_{ij}\propto1/|i-j|^{\alpha}$ with qubit separation number $|i-j|$, where in this work $\alpha \approx 1.1$.

\begin{figure}[t]
\vspace{0mm}
  \begin{center}
    \includegraphics[width=0.98\columnwidth]{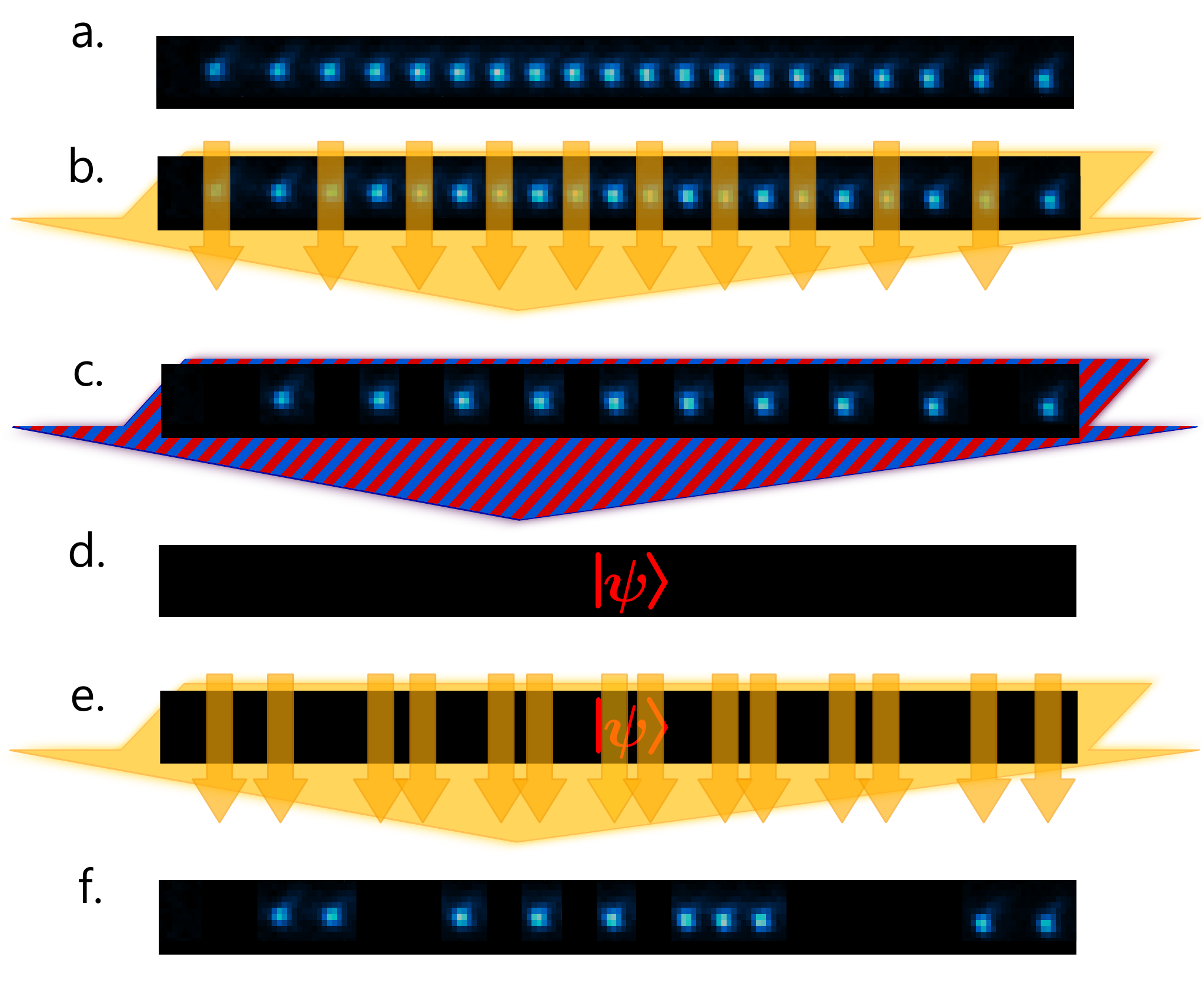}
   \vspace{-4mm}
   \caption{{\textbf{Conceptual schematic of the experiment}.
The experimental sequence proceeds as follows.
\textbf{a}. Standard Doppler cooling, optical pumping, and resolved-sideband cooling prepare the initial qubit state $\ket{0,0,0,...,0}$ and all 40 transverse vibrational string modes close ($<1$ phonon per mode) to the motional ground state~\cite{LanyonEtAl2017, JurcevicPHD}. An image of the 20 ions in our trap during state-dependent fluorescence detection of the state $\ket{0,0,0,...,0}$ is shown. The string length is 108 $\mu$m.
\textbf{b}. A combination of single-ion-focused and global laser beams prepares the initial state $\ket{\psi(t=0)}=\ket{1,0,1,...}$.
\textbf{c}.  A bichromatic light field is applied to $\ket{\psi(t=0)}$, subsequently inducing qubit-qubit interactions (not shown) as described in the text~\cite{LanyonEtAl2017, JurcevicPHD, KimChangIslamKorenblitDuanMonroe2009}.
\textbf{d}. After any desired evolution time, the interactions are turned off, leaving a nonclassical state of qubits (well approximated by) $\ket{\psi(t)}$.
\textbf{e}. A combination of single-ion-focused and global laser beams is used to rotate the basis of individual qubits, determining the desired measurement basis in the next step. An example laser pattern is shown.
\textbf{f}. The standard state-dependent resonant fluoresce technique is used to determine the state of each qubit (see Methods in Ref.~\cite{LanyonEtAl2017} for more details). An example outcome, imaged on an EMCCD camera, is shown schematically. The ions are then cooled again and initialised, ready for the sequence to be repeated.}
 }
   \label{figure0}
  \vspace{-8mm}
  \end{center}
\end{figure}

The ground state of $H_{XY}$ has all qubits in the state $\ket{0}$. The excited states are split into $m$ uncoupled and nondegenerate manifolds. Each manifold contains an integer number of qubit excitations (qubits in the state $\ket{1}$), with the $m$th manifold containing states with $m$ qubit excitations. In previous work, we have shown that an initial state consisting of a single localised qubit excitation coherently disperses in the system, distributing quantum correlations as it propagates~\cite{JurcevicEtAl2014}. Here, we study the entanglement generated during the time evolution of the initial 20-qubit N\'eel-ordered product state $\ket{\psi(t=0)}=\ket{1,0,1,...}$ under $H_{XY}$. That is, we study the state in the laboratory that is ideally described by $\ket{\psi(t)}=\exp(-i H_{XY}t)\ket{\psi(0)}$. The initial N\'eel state $\ket{\psi(0)}$ contains localised qubit excitations at every other site, which should ideally coherently disperse in the subsequent dynamics, entangling groups of neighbouring qubits, as we have previously shown for neighbouring pairs with a string of up to 14 qubits~\cite{LanyonEtAl2017}. While Ref.~\cite{LanyonEtAl2017} presented scalable tomography techniques~\cite{CramerPlenio2010, FlammiaGrossBartlettSomma2010, CramerPlenioFlammiaSommaGrossBartlettLandonCardinalPoulinLiu2010}, here we study multipartite entanglement dynamics in the system.

The initial state is prepared as follows. Standard Doppler cooling, optical pumping, and resolved-sideband cooling prepare the initial qubit state $\ket{0,0,0,...,0}$ and all 40 transverse vibrational string modes into the motional ground state~\cite{LanyonEtAl2017, JurcevicPHD}. Next, a combination of qubit-resonant laser beams that illuminate all ions simultaneously and off-resonant single-ion-focused laser beams flip every second qubit, preparing the state $\ket{\psi(0)}$. The interactions (laser-induced forces) simulating $H_{XY}$ are then turned on.

After the desired evolution time $t$, the interactions are turned off and the state [ideally $\ket{\psi(t)}$] is measured via qubit-state-dependent resonance fluorescence, using a single-ion-resolved electron multiplying charge coupled device (EMCCD) camera. Specifically, detecting a fluorescing (nonfluorescing) ion corresponds to the measurement outcome $\ket{0}$ ($\ket{1}$). Such a measurement corresponds to projecting each of the 20 qubits into the eigenstates of the Pauli $Z$ operator, for which there are $2^{20}$ possible outcomes each corresponding to a 20-qubit projective measurement outcome. After repeated state preparations and measurements in the ``$Z$ basis" any single-qubit expectation value ($\langle Z_i \rangle$), $2$-qubit correlator ($\langle Z_i Z_j \rangle$), or indeed any other $n$-qubit ``$Z$-type" correlator can be estimated between any qubits (up to $n=20$). Performing single-qubit operations, with a single-ion-focused laser, before the aforementioned measurement process enables projective measurement of any qubit in any desired basis, and therefore the construction of any multiqubit correlation function. That is, in this work full local control over the individual {qubits} is available and necessary for state preparation and analysis. {A conceptual schematic of our experimental protocol is presented in Fig.~\ref{figure0}}.

One approach to studying the entanglement properties of an $N$-qubit system is to perform full quantum state tomography to estimate the $N$-qubit density matrix and then develop and apply entanglement measures to that matrix. While this is technically possible for our 20-qubit system (i.e., the required measurements can each be performed in principle), it is practically not feasible as, e.g., billions of measurement bases are required. In general, the number of measurement bases required for full state tomography grows exponentially in $N$ as $3^N$. {Several} of us have recently shown that matrix product state (MPS) tomography can provide a pure-state estimate of states generated in quantum systems with finite-range interactions, using a number of measurements (and all other resources) that scales efficiently (polynomially) with the system size~\cite{LanyonEtAl2017}. However, MPS tomography failed to produce a useful pure-state description in our present 20-qubit system, probably due to errors in preparation of the initial state that lead to mixed states and the long-range nature of the interactions present in our 20-qubit Hamiltonian.

A more favourable approach to detecting and characterising entanglement in $N$-qubit systems is to develop entanglement witnesses that are not a function of every element of the density matrix and can be directly measured in the laboratory with a practical number of measurements.

\section{Initial Results: Magnetisation and Entanglement}\label{sec:initial results}

\begin{figure}[ht]
\vspace{0mm}
  \begin{center}
    \includegraphics[width=0.99\columnwidth]{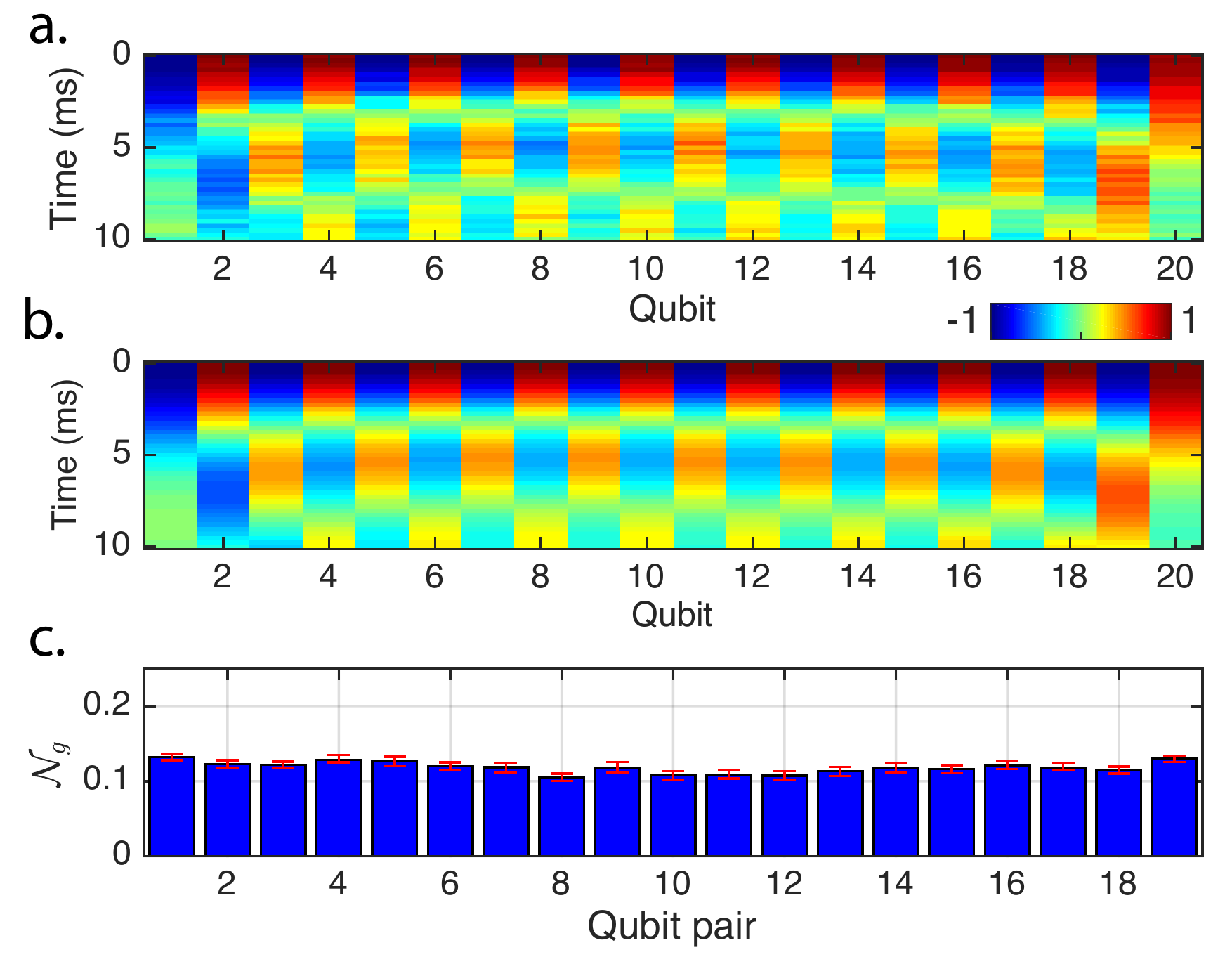}
   \vspace{-4mm}
   \caption{\textbf{Quantum dynamics of the 20-qubit register}.
 \textbf{a}. Experimental data. Single-qubit magnetisation $\langle Z_i \rangle$.
 \textbf{b}. Theoretical model. Single-qubit magnetisation $\langle Z_i \rangle$ for the exact time-evolved state $\ket{\psi(t)}$.
 \textbf{c}.  Experimental data. Entanglement in all 19 neighbouring qubit pairs at evolution time $t=2$ ms, quantified by the genuine multipartite negativity~\cite{JungnitschMoroderGuehne2011, HofmannMoroderGuehne2014}. Values for entanglement are calculated from experimentally reconstructed $2$-qubit density matrices. Errors are 1 s.d.\ derived from Monte Carlo simulation of finite measurement number.
 }
   \label{figure1}
  \vspace{-8mm}
  \end{center}
\end{figure}

As the first experimental step, we prepare the time-evolved state of our system [ideally described by $\ket{\psi(t)}$] and measure each qubit in the $Z$ basis. The dynamical evolution of $\langle Z_i \rangle$ (proportional to the probability of finding a qubit excitation at site $i$) shows how the multiple excitations in the initial state disperse, and then partially refocus at later times, in close agreement with predictions from the exact model (see Figs.~\ref{figure1}~a and~\ref{figure1}~b). One sees in the data and theory (Figs.~\ref{figure1}~a and~\ref{figure1}~b) that the qubit excitations at the ends of the string disperse more slowly than those in the centre of the string (the green color representing ``delocalised" excitations appears at later times for qubits farther from the centre of the string). The qubit-qubit interactions ($J_{ij}$) are not homogeneous across the string, leading to slower evolution for qubits farther away from the centre. $J_{ij}$ for spin pairs near the string centre is approximately 25\% larger than for spin pairs located at either end of the string. The physical origin of this effect is a combination of the Gaussian intensity profile of the laser fields involved in generating the interactions (the laser beam is centred on the middle ions and therefore weaker on the outer ions)  and the intrinsic interaction inhomogeneity across the string set by the ion-string vibrational mode frequencies, for our chosen trapping parameters. Both effects are taken into account in the theoretical model presented in Fig.~\ref{figure1}~b.

As the final experimental step, we perform the set of measurements that would be sufficient to reconstruct (via full quantum state tomography) the density matrices of all neighbouring $k=3$ qubits (qubit triplets), during the simulator dynamics. This set consists of $3^k=27$ measurement bases (with 1000 measurements performed per basis), corresponding to all possible combinations of choosing three Pauli operators. We carry out a simple scheme (choice of measured Pauli operators) that allows measurements on all 18 neighbouring qubit triplets (out of the 20-qubit string) to be performed in parallel, requiring a total of only twenty-seven 20-qubit measurement bases. From this data set, we could reconstruct the density matrices of all single qubits, neighbouring qubit pairs, and neighbouring qubit triplets. Generalising this approach to arbitrary $k$, all $N-k+1$ groups of neighbouring $k$-qubit density matrices in an $N$-qubit string can be fully characterised by measuring in $3^k$ bases (independent of the number of qubits $N$). For fixed $k$, this measurement approach is clearly efficient (constant overhead) in the system size $N$. We nonetheless stop at $k=3$, as the number of measurement bases for $k=4$ is already quite demanding and, as we show, {$k=3$} is already sufficient to observe genuine multipartite entanglement in groups of up to five qubits.

We reconstruct the density matrices of all neighbouring qubit pairs from the experimental data, via the standard maximum likelihood method, which finds the most likely physical density matrix to have produced the data. For each of the reconstructed $2$-qubit states we evaluate the genuine multipartite negativity $\mathcal{N}_g$, an established measure for GME~\cite{JungnitschMoroderGuehne2011, HofmannMoroderGuehne2014}. A positive value of $\mathcal{N}_g$  for a given $k$-qubit state implies the existence of genuine $k$-partite entanglement in this state, since $\mathcal{N}_g$ vanishes for all biseparable states. For two qubits, $\mathcal{N}_g$ is directly related to the logarithmic negativity~\cite{Plenio2005}. More details on $\mathcal{N}_g$ are given in Sec.~\ref{sec:Ulm witnesses main text}. From the results one sees that all neighbouring qubit pairs become entangled during the time evolution of the system, as is shown for $t=2$~ms in Fig.~\ref{figure1}~c. Error bars, on properties calculated using the tomographically reconstructed density matrices, are derived from the finite number of measurements (1000 for each global basis) used to estimate expectation values and calculated using the standard {Monte Carlo method~\cite{EfronTibshirani1986}}.

Naturally, one may wonder if entanglement extends beyond qubit pairs, for instance, in the form of bipartite entanglement between distant qubits or in terms of genuine multipartite entanglement between groups of more than two (adjacent) qubits. In fact, one may even be tempted to ask, \emph{is multipartite entanglement not implied if every neighbouring qubit pair is entangled?}
The answer to this question is simply no. One can indeed have states that feature entanglement in every $2$-qubit reduction, yet still feature only bipartite entanglement%
\footnote{Take, for example, the $k$-qubit state with density operator $\rho=\frac{1}{2}(\bigotimes_{i=1}^{k/2}\ket{\phi_{2i-1,2i}^+}\!\!\bra{\phi_{2i-1,2i}^+}+\bigotimes_{i=1}^{k/2}\ket{\phi_{2i,2i+1}^+}\!\!\bra{\phi_{2i,2i+1}^+})$ where  $k$ is assumed to be even and $\ket{\phi_{i,j}^+}=\tfrac{1}{\sqrt{2}}(\ket{0}_{i}\ket{0}_{j}+\ket{1}_{i}\ket{1}_{j})$ with the second index being understood as modulo $k$. The state $\rho$ is a convex combination of $\frac{k}{2}$-separable states, yet the reduced state $\rho_{i,i+1}=\frac{1}{2}(\ket{\phi^+}\!\!\bra{\phi^+}+\tfrac{1}{4}\id)$ of every pair of neighbouring qubits is entangled.}. Nonetheless, in theory, it is often possible to detect GME purely from inspection of the reduced density matrices of overlapping groups of qubits~\cite{GuehneToth2009}. This is the basis for our first approach to detecting GME, presented in the next section, where we derive GME witnesses purely from neighbouring two-body observables.

In general, the task of determining if and how an $N$-qubit quantum system is entangled is highly nontrivial. For an arbitrary \emph{known} mixed state (e.g., reconstructed via full state tomography), the problem is at least NP hard and even the best known relaxations are semidefinite programs (SDPs) that are not feasible beyond 5 qubits with our computers~\cite{JungnitschMoroderGuehne2011, LancienGuehneSenguptaHuber2015}. However, if the density matrix is close to a given pure target state $\ket{\psi_T}$, then targeted witnesses can be constructed to detect its entanglement without resorting to full tomography~\cite{GuehneToth2009}. There, the canonical ansatz would be to estimate the fidelity to the ideal target state $\tr(\rho\ket{\psi_T}\!\!\bra{\psi_T})$ and check whether it is above the maximum possible fidelity of a biseparable state. That is, a corresponding witness would be $W=\beta\id-\ket{\psi_T}\!\!\bra{\psi_T}$, where $\beta:=\min_{A|\overline{A}}||\tr_{\overline{A}}(\ket{\psi_T}\!\!\bra{\psi_T})||_\infty$, see, e.g., Ref.~\cite[Sec.~3.6]{GuehneToth2009}. While this witness could in principle be successful in detecting GME if the experimental state is indeed very close to the intended pure state, it suffers from poor noise resistance\footnote{For qubits $\beta\leq\frac{1}{2}$ and thus the best possible noise resistance is $50\%$ white noise. While this may seem strong from a bipartite intuition, many multipartite states in fact have asymptotically perfect white noise resistance, i.e., approaching $100\%$ polynomially in system dimension, which would inevitably be missed by such simple witness constructions.} and the task of determining the state fidelity for arbitrary pure states still requires {a number of measurement settings that scales exponentially in qubit number}~\cite{FlammiaLiu2011}.

For example, if the state is ``well conditioned''~\cite{FlammiaLiu2011}, i.e., if only few Pauli-expectation values are of a significant size and all others vanish, one could estimate the fidelity via randomised measurements~\cite{FlammiaLiu2011} {with effort that scales efficiently in qubit number}. Although our states are not well conditioned, in Ref.~\cite{LanyonEtAl2017} we implemented this randomised measurement strategy to obtain a fidelity estimate for a 14-qubit version of the states presented here, at one time step. {That experiment required preparing 5$\times10^5$ sequential copies of the state and involved over 5 hours of data taking (with periodic recalibration of experimental parameters). Numerical simulations of the randomised measurement technique applied to the 20-qubit states considered in this work show that more than 3 times the
number of copies, and therefore impractical measurement time, would be required to yield accurate fidelity estimations. As such, we aim to develop novel, and more time-efficient, approaches to characterising entanglement in our 20-qubit system. As a first step in this direction, we focus on the dynamics of subsystem entanglement percolating through the system and are able to make statements about the entanglement using measurements completed in a few tens of minutes in our system.}

\section{GME witnesses based on $2$-qubit observables}\label{sec:Vienna witnesses main text}

In this section, we construct analytical witnesses for GME based on $2$-qubit observables and use them to detect GME in groups of up to three neighbouring qubits ($k=3$) within the $20$-qubit ($N=20$) register. Recall that a (multipartite) pure state $\ket{\psi}$ is called biseparable if there exists a bipartition $A|B$ such that $\ket{\psi}=\ket{\phi}_{A}\ket{\chi}_{B}$ for some $\ket{\phi}_{A}$ and $\ket{\chi}_{B}$, and is called genuinely multipartite entangled otherwise. Mixed states are GME if their density operators cannot be written as convex combinations of biseparable pure states. For more details, see Appendix~\hyperref[sec:GME witnesses]{A.II}.

Following the observation in the previous section of strong entanglement between neighbouring qubits, the first type of GME witnesses we consider is based on average fidelities of the $2$-qubit density matrices with Bell states. As such, only expectation values of pairs of Pauli operators, on $k$-qubit subsets of choice, are required. Linear combinations of these expectation values are then evaluated and compared to their respective thresholds for biseparable $k$-qubit states. Surpassing a $k$-qubit biseparability threshold then detects genuine $k$-partite entanglement.

We now present a short technical summary of our method, and refer the reader to Appendix~\ref{appendix:vienna witnesses} for more details. The main quantity of interest for detecting $k$-qubit GME in this section is the $k$-qubit \emph{symmetric average Bell fidelity} $\bar{\mathcal{F}}^{(k)}_{\mathrm{Bell}}$, which we define as
\begin{align}
    \bar{\mathcal{F}}^{(k)}_{\mathrm{Bell}}& :=
    \tfrac{1}{4b_{k}}\Bigl(b_{k}\!+\!\!\sum\limits_{\substack{i,j=1 \\ i<j}}^{k}
    \bigl(|\!\expval{\tilde{X}_{i}\tilde{X}_{j}}\!|\!+\!|\!\expval{\tilde{Y}_{i}\tilde{Y}_{j}}\!|\!+\!|\!\expval{\tilde{Z}_{i}\tilde{Z}_{j}}\!|\bigr)
    \Bigr),
    \label{eq:NQB symm average Bell fid intro}
\end{align}
where $b_{k}=\tbinom{k}{2}=\tfrac{1}{2}\tfrac{k!}{(k-2)!}$, and the subscripts $i$ and $j$ denote operators acting nontrivially only on the $i$th and $j$th qubits, i.e., $O_{i}   \equiv \mathds{1}_{1}\otimes\ldots\otimes\mathds{1}_{i-1}\otimes O_{i}\otimes\mathds{1}_{i+1}\otimes\ldots\otimes\mathds{1}_{N}$. The operator triple $\tilde{X}_{i}=U_{i}XU_{i}^{\dagger}$, $\tilde{Y}_{i}=U_{i}YU_{i}^{\dagger}$, and $\tilde{Z}_{i}=U_{i}ZU_{i}^{\dagger}$ is chosen unitarily equivalent to the usual triple of Pauli operators $X$, $Y$, and $Z$, although the unitary $U_{i}\in SU(2)$ may be chosen differently for each qubit (for each $i$). This ensures that $\bar{\mathcal{F}}^{(k)}_{\mathrm{Bell}}$ can be written as a linear combination (the absolute values can be replaced by appropriate sign changes) of pairs of Pauli operators. As we show in detail in Appendix~\ref{appendix:vienna witnesses}, any quantum state of $k$ qubits for which
\begin{align}
     \bar{\mathcal{F}}^{(k)}_{\mathrm{Bell}} &>
    \begin{cases}
        \tfrac{1}{12}\bigl(3+\sqrt{15}\bigr)    &   \mbox{for $k=3$}\\[1mm]
        \tfrac{1}{4}\bigl(1+\sqrt{3}\bigr)-\tfrac{1}{2k}\bigl(\sqrt{3}-1\bigr)  & \mbox{for $k\geq 4$}
    \end{cases}
    \label{eq:NQB symm average Bell fid witness}
\end{align}
is genuinely $k$-partite entangled for any choice of $U_{1}, \ldots, U_{k}$. For example, for $k=3$ and $k=4$ one can detect GME for $\bar{\mathcal{F}}^{(3)}_{\mathrm{Bell}}>\tfrac{1}{12}\bigl(3+\sqrt{15}\bigr)\approx0.573$ and for $\bar{\mathcal{F}}^{(4)}_{\mathrm{Bell}}>\tfrac{1}{8}\bigl(3+\sqrt{3}\bigr)\approx0.592$, respectively. {Meanwhile}, the threshold for $k=2$ qubits, $\bar{\mathcal{F}}^{(2)}_{\mathrm{Bell}}>0.5$, is a well-known result.

If the underlying $N$-partite quantum state is known (the $U_i$ are known), one could  directly measure all of the $3b_{k}$ $2$-qubit correlators $\expval{\tilde{O}_{i}\tilde{O}_{j}}$ appearing in Eq.~(\ref{eq:NQB symm average Bell fid intro}) for optimally chosen $\{U_{i}\}$. However, when the optimal local measurements are unknown, one strategy is to measure the $6k$ $k$-qubit basis settings corresponding to the set
\begin{align}
    \{O\suptiny{0}{0}{(i)}_{XY},O\suptiny{0}{0}{(i)}_{YX},O\suptiny{0}{0}{(i)}_{XZ},O\suptiny{0}{0}{(i)}_{ZX},O\suptiny{0}{0}{(i)}_{YZ},O\suptiny{0}{0}{(i)}_{ZY}\}_{i=1,\ldots,k},
\end{align}
where $O\suptiny{0}{0}{(i)}_{AB}=A_{i}\prod_{i\neq j}B_{j}$, and perform the optimisation when evaluating the corresponding results. From the $2^{k}$ outcomes of each of these simultaneous measurements of $k$ qubits one can obtain the expectation values of all pairwise combinations of Pauli operators.

\begin{figure*}[ht!]
    \begin{center}
        \includegraphics[width=1\textwidth,trim={0cm 0cm 0cm 0cm},clip]{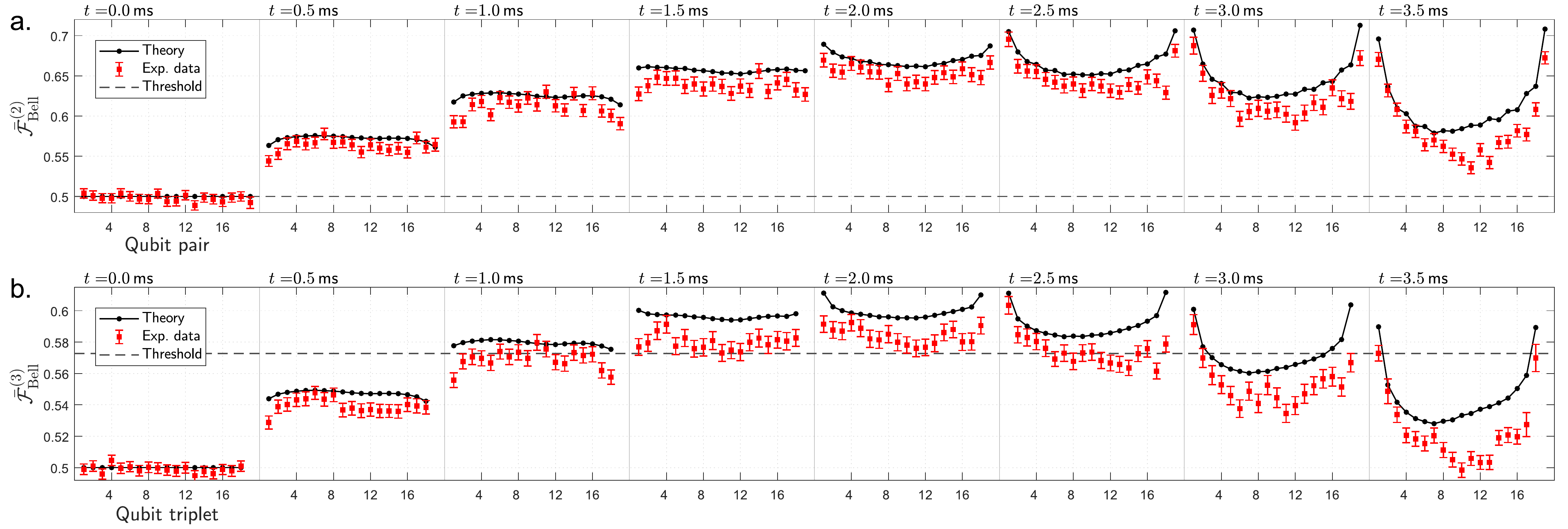}
        \vspace{-4mm}
        \caption{\textbf{Entanglement witnesses based on symmetric average Bell fidelities}. The experimental results (red) and theoretical predictions based on ideal time-evolved state $\ket{\psi(t)}$ (blue) for the entanglement witnesses $\bar{\mathcal{F}}^{(2)}_{\mathrm{Bell}}$ and $\bar{\mathcal{F}}^{(3)}_{\mathrm{Bell}}$ are shown in \textbf{a} and \textbf{b}, respectively. The horizontally arranged panels show the results at different time steps, 0.0 ms, 0.5 ms, 1.0 ms, etc., with intervals of 0.5~ms during the time evolution of the $20$-qubit chain, starting with 0~ms (the initial state). Within each panel, each dot represents a pair (\textbf{a}) or triple (\textbf{b}) of neighbouring qubits. The horizontal dashed lines indicate the detection thresholds for bipartite (\textbf{a}, $\bar{\mathcal{F}}^{(2)}_{\mathrm{Bell}}>0.5$) and genuine tripartite entanglement (\textbf{b}, $\bar{\mathcal{F}}^{(3)}_{\mathrm{Bell}}>\tfrac{1}{12}(3+\sqrt{15})\approx0.573$), respectively. The {error bars} for each qubit pair or triple represent 1 standard deviation of the mean in each direction. It can be seen in \textbf{a} that all qubits immediately become entangled with their direct neighbours and remain entangled throughout, whereas genuine tripartite entanglement is detected in time step 1.0 ms for the first time. At time step 2.0 ms, the witness $\bar{\mathcal{F}}^{(3)}_{\mathrm{Bell}}$ indicates that all neighbouring qubit triples are genuinely tripartite entangled simultaneously (although the witness is less than 1 standard deviation above the threshold for two of these triples).
        }
   \label{figure2}
  \vspace{-4mm}
  \end{center}
\end{figure*}

For our purposes, we exploit the fact that the results from the twenty-seven $20$-qubit measurement bases already taken in the laboratory are also sufficient to calculate all the expectation values appearing in the witnesses $\bar{\mathcal{F}}^{(k)}_{\mathrm{Bell}}$ for $k=2$ and $k=3$. That is, they contain as a subset, all the 2-qubit observables required to calculate $\bar{\mathcal{F}}^{(2)}_{\mathrm{Bell}}$ and $\bar{\mathcal{F}}^{(3)}_{\mathrm{Bell}}$, without knowledge of the states. The results, for increasing system interaction times and for the optimisation of the $U_{i}$ limited to the $X$-$Y$ plane for each qubit, are shown in Fig.~\ref{figure2}. First, the witness for bipartite entanglement ($\bar{\mathcal{F}}^{(2)}_{\mathrm{Bell}}>0.5$) reaffirms that all qubits are (bipartite) entangled with their direct neighbours throughout the interaction time (from time 0.5 to 3.5~ms). Second, genuine tripartite entanglement between neighbouring qubit triples  builds up more slowly and is initially detected at time 1.5~ms. At time 2~ms, {most triples of neighbouring qubits are genuinely tripartite entangled,} before the GME gradually disappears again at later times.

The experimental uncertainties (error bars) in Fig.~\ref{figure2} originate from a finite number of measurements used to estimate expectation values. Specifically, the error bars show an estimate of 1 standard deviation of the mean.
When estimating the standard deviation of the mean, one must consider possible correlations between $2$-qubit expectation values if they are estimated from outcomes of the same 20-qubit measurement basis. We estimate the relevant variance as described in Ref.~\cite[Supplementary Information Sections IV.A.4 and IV.A.5 on pages 9--13]{LanyonEtAl2017}.

The small deviations between the theoretical and measured dynamics of $\bar{\mathcal{F}}^{(2)}_{\mathrm{Bell}}$ and $\bar{\mathcal{F}}^{(3)}_{\mathrm{Bell}}$ in Fig.~\ref{figure2} are due to experimental imperfections, which we discuss in the next section.

A number of interesting observations, based on analyses beyond those presented in Fig.~\ref{figure2}, are now made. First, up to the evolution time presented in Fig.~\ref{figure2}, entanglement between any $2$-qubits spaced farther apart than direct neighbours was never detected\textemdash in agreement with the ideal theoretical model. For instance, on these timescales, qubit 1 does not directly become entangled with qubit 3 alone, but qubits 1, 2, and 3 do become genuinely tripartite entangled with each other. In fact, the absence of next-nearest neighbour pairwise entanglement was necessary to detect 3-qubit GME. Specifically, we found that a GME witness based only on the entanglement between direct neighbours (see Appendix~\hyperref[sec:Nearest-Neighbour Average Bell Fidelity as GME Witness]{A.II.2}) is not able to verify genuine tripartite entanglement, and it was only possible to do so once the (separable) correlations between non-neighbouring qubits (e.g., qubits 1 and 3) are also taken into account.

Second, although there are states for which the witness of Eqs.~(\ref{eq:NQB symm average Bell fid intro}) and~(\ref{eq:NQB symm average Bell fid witness}) could be used to detect GME between more than 3 parties\footnote{For instance, pure $4$-qubit Dicke states with two excitations would yield $\bar{\mathcal{F}}^{(4)}_{\mathrm{Bell}}=\tfrac{2}{3}>0.592$. Beyond $4$ qubits, we cannot say with certainty whether states exist for which our witness could certify GME in principle. For a discussion see Appendix~\hyperref[sec:Symmetric Average Bell Fidelity as GME Witness]{A.II.3}}, it is not sensitive enough to do so for the states presented here in our setup (neither for the theoretical predictions nor for the experimental data).

To address the question of whether genuine multipartite quantum correlations occur in groups of more than 3~qubits, in our setup, we hence turn to more computationally demanding procedures, which we present in the next section.

{The observed and predicted entanglement peak amplitude and dynamics for qubits near the centre and those near the ends (Fig.~\ref{figure2}) are markedly different. We attribute those differences to the interaction inhomogeneity across the qubit string and boundary effects.}


\section{GME Witnesses based on numerical search}\label{sec:Ulm witnesses main text}

In this section, we present and apply a method that employs a numerical search to find $k$-qubit witnesses for GME. This search is computationally intensive: an optimisation is performed that takes computational resources that increase exponentially with $k$. Finding a GME witness operator for mixed $5$-qubit states is already at the practical limit of our available computers and algorithms. Nonetheless, we find witnesses that succeed in detecting GME in groups of up to 5 qubits in our 20-qubit experimental system. In the following, we give a brief overview of the new witnesses and defer to Appendix~\hyperref[appendix:ulm witnesses]{B} for a more detailed discussion of the technical aspects.

We make use of the genuine multipartite negativity ($\mathcal N_{\rm g}$), an established measure for GME~\cite{JungnitschMoroderGuehne2011, HofmannMoroderGuehne2014}. A positive value of $\mathcal N_{\rm g}$ for a given $k$-qubit state implies the existence of genuine $k$-partite entanglement in this state, since $\mathcal N_{\rm g}$ vanishes for all biseparable states. The $\mathcal N_{\rm g}$ can be calculated given knowledge of the density matrix~\cite{JungnitschMoroderGuehne2011, HofmannMoroderGuehne2014}. However, we have not performed a tomographically complete set of measurements for more than $k=3$ qubits (we do not have the density matrices for the state in the lab, for $k>3$). Our approach, to detect GME in any given group $i$, of $k$~qubits, is to find a $k$-qubit witness operator $Q^{(k)}_i$ whose expectation value provides a lower bound on the $k$-qubit $\mathcal N_{\rm g}$, and which can be written as a function of the set of measurements that were carried out in our experiment. We now provide more details on this approach.

\begin{figure*}[ht]
  \begin{center}
        \includegraphics[width=1\textwidth,trim={0cm 0cm 0cm 0cm},clip]{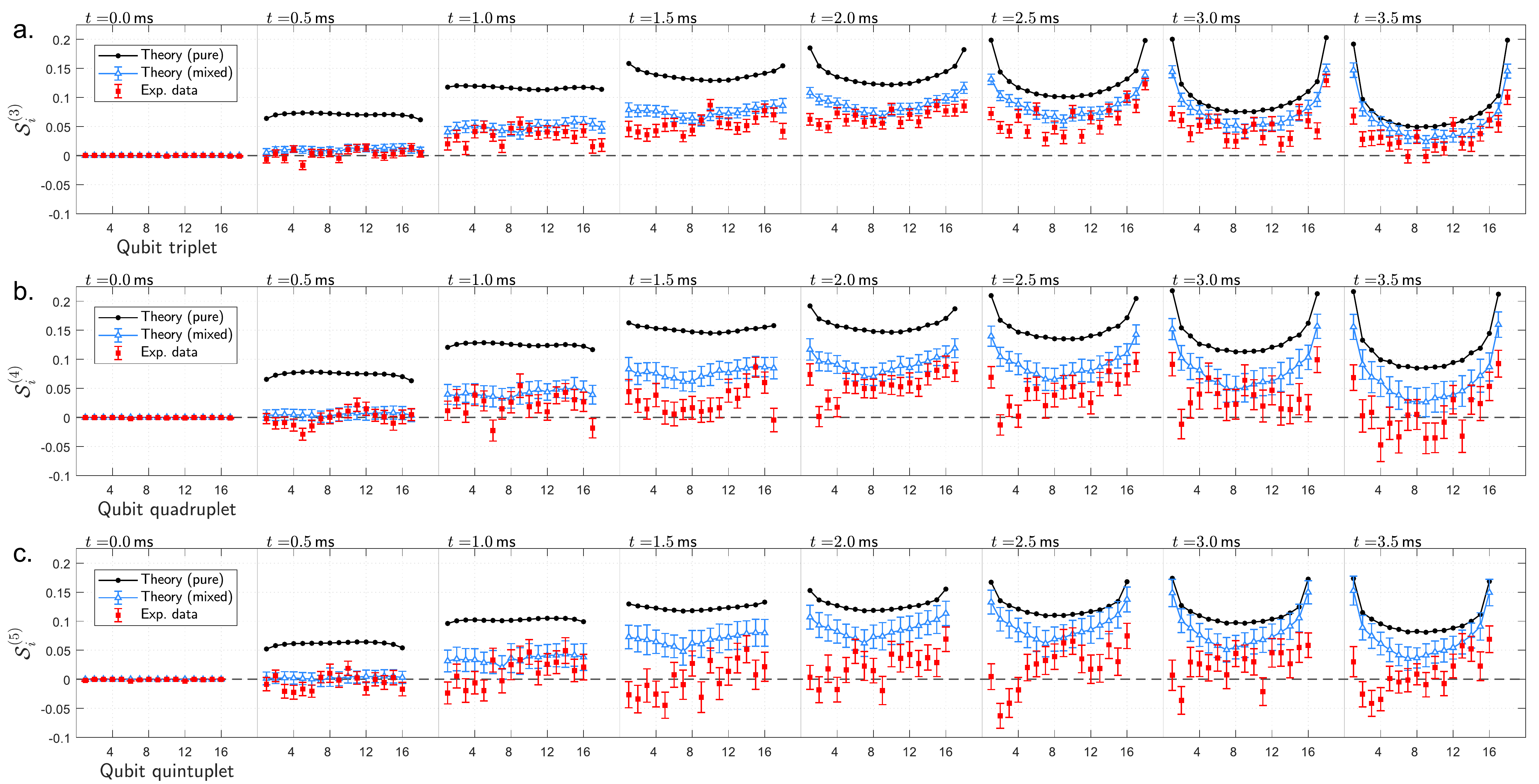}
   \vspace{-6mm}
   \caption{\textbf{Entanglement witnesses derived by a numerical search}. The entanglement witnesses $\mathcal{S}_i^{(3)}$, $\mathcal{S}_i^{(4)}$, and $\mathcal{S}_i^{(5)}$ (see Sec.~\ref{sec:Ulm witnesses main text}) are shown in \textbf{a}-\textbf{c}, respectively. The horizontally arranged subpanels show the results at different system evolution time steps with intervals of 0.5~ms during the time evolution of the $20$-qubit chain, starting with 0~ms (the initial state). Within each panel, each dot represents results for a given triplet (\textbf{a}, $k=3$), quadruplet (\textbf{b}, $k=4$) or quintuplet (\textbf{c}, $k=5$) of neighbouring qubits. {GME is detected if the witness is positive.}
   The different theory plots are for the pure model (black circles without error bars) and mixed model (blue {triangles} with error bars), as described in the text. The red squares with error bars show experimental data. All error bars show 1 standard deviation of the mean and originate from a finite number of numerically simulated measurements per measurement basis (1000).
    }
   \label{figure3}
  \vspace{-5mm}
  \end{center}
 \end{figure*}

We perform a search to find a $k$-qubit operator $Q^{(k)}_i$, subject to two important constraints. First, we search for an operator which both maximises the following inequality,
\begin{equation}
\label{QEW}
-\tr(Q^{(k)}_i \rho^k_i) \equiv \mathcal{S}_i^{(k)} \le \mathcal{N}_{\rm g}({\rho^k_i})
\end{equation}
for a specific $k$-qubit state of interest $\rho^k_i$, and satisfies the inequality for all possible $k$-qubit states.
We call $Q^{(k)}_i$ a quantitative entanglement witness (QEW) because it provides a lower bound on $\mathcal N_g$.
It is straightforward to constrain the search in this way and also to verify that any given $Q^{(k)}_i$ is a QEW. When searching for the optimal witness, we use a theoretical model for the time-evolved $k$-qubit state for $\rho^k_i$.
Second, we include the additional constraint that the $Q^{(k)}_i$ can be written as a linear function of the $k$-qubit measurement operators (projectors) that were done, involving qubit group $i$. Specifically, we restrict $Q^{(k)}_i$ to the form
\begin{equation}
\label{projector QEW}
Q^{(k)}_i = \sum_{\vec{s},\vec{\alpha}}  c^{(k)}_{i;\vec{s},\vec{\alpha}} P_{\vec{s},\vec{\alpha}}^{(k)},
\vspace*{-2mm}
\end{equation}
where the projectors $P_{\vec{s},\vec{\alpha}}^{(k)}$ correspond to the marginal distributions of the twenty-seven 20-qubit projective measurement settings carried out in the lab (Sec.~\ref{sec:initial results}), and $c^{(k)}_{i;\vec{s},\vec{\alpha}}$ denote some coefficients. Here, $\vec{s}$ and $\vec{\alpha}$ label, respectively, the qubit outcome and the local basis of the measurements, see also Appendix~\hyperref[sec:27 Accessible measurements]{B.II}.

The search for the optimal witnesses operator $Q^{(k)}_i$ is carried out using a semidefinite program. The run time of the SDP is polynomial in the dimension of the Hilbert space~\cite{BoydVandenbergheBook2004, TutuncuTohTodd2003} but the dimension of our Hilbert space naturally increases exponentially with the number of qubits $k$. This makes the optimisation demanding already for medium numbers of qubits: Our available computational resources are not sufficient to determine optimal witnesses for states of more than 5 qubits.
The $Q^{(k)}_i$ which satisfies Eqs.~\eqref{QEW} and~\eqref{projector QEW}, and maximises the left-hand side of Eq.~\eqref{QEW}, determines an optimal witness tailored to the target state (from a theoretical model) and the available measurements. Once this optimal $Q^{(k)}_i$ is found we can calculate its expectation value from the outcomes of the measurement done on the state in the laboratory. A witness expectation value ($\mathcal{S}_i^{(k)}$) larger than zero then detects $k$-qubit GME ($\mathcal N_{\rm g}>0$), for the $i$th group of $k$~qubits.

The experimental results for $k=3$ {presented in Fig.~\ref{figure3}~a} show that all neighbouring qubit triplets soon develop GME during the dynamics, to within many standard deviations, reaching a maximum at $t=2$~ms. Furthermore, for the times 2, 2.5, and 3~ms, {Figs.~\ref{figure3}~b and~\ref{figure3}~c show that} GME is detected in the majority of all neighbouring groups of 4 and 5 qubits, to within at least 1 standard deviation of experimental uncertainty.

Figure~\ref{figure3} compares the witness results obtained from the data with those derived from two theoretical models. The first ``pure" model employs the perfect pure 20-qubit time-evolved state [$\ket{\psi(t)}$] and uses exact knowledge of $k$-qubit density matrices to optimise and apply the witnesses. {The} witness expectation value for the pure model yields $\mathcal{S}_i^{(k)}=\mathcal{N}_{\rm g}({\rho^k_i})$. Although the pure model succeeds in qualitatively describing the multipartite entanglement dynamics, the witness expectation values from the data are generally offset to lower values.
A more sophisticated ``mixed" model is able to explain part of this offset, which includes known imperfections in preparing the initial N\'eel-ordered state. Specifically, out of 1000 attempts to generate the N\'eel state, we observe the correct output state 829 times. In the remaining 171 cases, 146 correspond to single qubit flip errors and the rest to errors with two or more qubit flips.
We attribute these errors to uncontrolled fluctuations in laser intensity and frequency, and model them as leading to the preparation of a statistical mixture of those different logical initial states, with corresponding weights.
The witnesses used for the data were obtained by a search involving the mixed-model reduced states for the targets. We attribute the remaining small differences between data and theory, in Fig.~\ref{figure3}, to additional mixing processes that occur in the laser-induced qubit-qubit interactions.

The mixed theory predictions in Fig.~\ref{figure3} include error bars due to the use of a finite number of numerically simulated measurements per measurement basis (1000).
Error bars indicate 1 standard deviation of the mean, which is estimated as in Sec.~\ref{sec:Vienna witnesses main text}, and show that the fluctuations in the data are largely consistent with those expected from such statistical noise.
We conclude that, in order to witness 4- and 5-partite GME with greater statistical significance in future work, we could benefit from taking more measurements. However, it will be challenging to ensure that the experimental configuration remains stable over the longer time required to take such additional measurements.

{The} sizes of the error bars on both data and mixed theory points, in Fig.~\ref{figure3}, increase with increasing $k$. This can be understood as follows: there are more measurement outcomes available in the data for the $k=3$ witness calculations than for larger $k$. Amongst the twenty-seven 20-qubit measurement bases, 3-qubit measurements are repeated (duplicated) more often in the measurement pattern than 4-qubit or 5-qubit measurements, leading to better statistics.

\vspace*{-4mm}
\section{Discussion and Conclusion}\label{sec:discussion}
\vspace*{-3mm}

We have experimentally generated and detected the presence of entanglement in a register of 20 qubits. In particular, we detected the dynamical evolution of genuine multipartite entanglement in the system following a quench, and developed new characterisation techniques to do so. While we cannot say that 20-qubit GME was generated, we can say that every qubit simultaneously became genuine multipartite entangled with a least two of its neighbours and, in most cases, three and four of its neighbours.

Our experimental apparatus represents the largest joint system of individually controllable subsystems to date where the presence of entanglement has been demonstrated. Each qubit can be individually controlled and qubit-qubit interactions can be turned on and off as desired (and tuned to have various forms). As such, our system has the capability to perform universal quantum simulation and quantum computation.

Confirming GME beyond groups of 5 qubits, even for the ideal states, is currently beyond our available classical computational resources and algorithms. A possible approach to overcome that problem is to exploit symmetries in the system and initial state, to reduce the size of the search space for witnesses. Another is to tune the experimental system Hamiltonian into regimes where more symmetries are apparent or approximated, e.g., infinite or nearest-neighbour-only qubit-qubit interaction ranges.

Finally, witnesses based on average Bell-state fidelities are straightforward to use and measure in the lab. As with all such witnesses, they detect entanglement without the need to carry out state tomography and can be evaluated with only a few measurements. This can be important for the detection of weakly entangled states, where estimates based on state tomography are known to overestimate entanglement~\cite{SchwemmerKnipsRichartWeinfurterMoroderKleinmannGuehne2013}. Our witnesses based on brute-force numerical searching have the advantage of placing the least constraints on the form of the state in the lab and the measurements that should be taken: this can be important in the case of unknown local rotations of qubits during the dynamics. As such, our witnesses should find application beyond the present trapped-ion setting.

\vspace*{-4mm}
\begin{acknowledgments}
\vspace*{-3mm}
The work in Innsbruck was supported by the Austrian Science Fund (FWF) under Grant No. P25354-N20, by the European Commission via the integrated project SIQS, and by the Army Research Laboratory under Cooperative Agreement No. W911NF-15-2-0060.  The views and conclusions contained in this document are those of the authors and should not be interpreted as representing the official policies, either expressed or implied, of the Army Research Laboratory or the U.S. Government. B.~L. acknowledges funding from the Austrian Science Fund (FWF) through the  START  project Y 849-N20. N.~F. and M.~Huber acknowledge funding from the Austrian Science Fund (FWF) through the START project Y879-N27 and the joint Czech-Austrian project MultiQUEST (I 3053-N27). The work in Ulm was supported by an Alexander von Humboldt Professorship, the ERC Synergy grant BioQ, the EU projects QUCHIP, and the U.S. Army Research Office Grant No. W91-1NF-14-1-0133, as well as by computational resources by the state of Baden-W{\"u}rttemberg through bwHPC and the German Research Foundation (DFG) through Grant No. INST 40/467-1 FUGG.
\end{acknowledgments}


\bibliography{bibfile}



\hypertarget{sec:appendix}
\appendix
\section*{Appendices}
\renewcommand{\thesubsubsection}{A.\Roman{subsection}.\arabic{subsubsection}}
\renewcommand{\thesubsection}{A.\Roman{subsection}}
\renewcommand{\thesection}{A}
\setcounter{equation}{0}
\numberwithin{equation}{section}
\setcounter{figure}{0}
\renewcommand{\thefigure}{A.\arabic{figure}}

\section{GME Witnesses Based on Bipartite Correlators}\label{appendix:vienna witnesses}

In this section of the appendix, we introduce a method for the detection of \emph{genuine multipartite entanglement} in $N$-qubit systems. This method is based on $2$-qubit observables and does not require full state tomography. In particular, our detection criteria can be phrased as biseparability thresholds for average Bell fidelities, i.e., expectation values of linear combinations of pairs of Pauli operators. At the heart of this method lies the \emph{anticommutativity theorem} (ACT) from Refs.~\cite{TothGuehne2005, AsadianErkerHuberKlockl2016}, which we use to provide bounds on the average Bell fidelities. Although our approach is not able to detect all types of GME in multipartite systems, its advantage lies in providing linear entanglement witnesses that can be practically evaluated with only a few measurements. More specifically, our approach does not require obtaining a good estimate of the $N$-partite correlation tensor~\cite{WuKampermannBrussKloecklHuber2012, KloecklHuber2015} with $3^{N}$ components, but instead only needs at most $6N$ measurements of strings of $N$ local Pauli operators to test for $N$-partite GME. As we discuss, the linearity of the witness also makes it amenable to a simple treatment of the potentially correlated statistical errors arising from deriving expectation values of bipartite observables from simultaneous measurements of $N$ qubits.

Following the brief description of these results in Sec.~\ref{sec:Vienna witnesses main text} of the main text, we now present more detailed derivations of the quantities and bounds that we consider. In Appendix~\hyperref[sec:framework]{A.I} we briefly define and motivate the basic quantities of interest, before we construct our GME witnesses in Appendix~\hyperref[sec:framework]{A.II}.


\vspace*{-3mm}
\subsection{Framework}\label{sec:framework}

In this section, we explain the basic quantities and notions of interest, i.e., the anticommutativity theorem of Ref.~\cite{AsadianErkerHuberKlockl2016} and the average Bell fidelities to establish a basis for the more detailed discussion of GME that is to follow in Appendix~\hyperref[sec:GME witnesses]{A.II}.


\vspace*{-3mm}
\subsubsection{The Anticommutativity Theorem}\label{sec:ACT theorem}

Let us consider a set $\{A_{n}\}_{n=1,2,\ldots,k}$ of self-adjoint, normalized, anticommuting operators on a Hilbert space $\mathcal{H}$ with $\dim(\mathcal{H})=d$, i.e.,
\begin{align}
    \tr\{A_{m},A_{n}\}_{+}  &=\tr\bigl(A_{m}A_{n}+A_{n}A_{m}\bigr)=2d\delta_{mn}
    \label{eq: normalized anticomm operators}
\end{align}
for all $m,n=1,\ldots,k$. The \emph{anticommutativity theorem}~\cite{TothGuehne2005, AsadianErkerHuberKlockl2016} then states that for all states $\rho\in L(\mathcal{H})$,
\begin{align}
    \sum\limits_{n=1}^{k}\expval{A_{n}}^{2}_{\rho} &\leq\,\max_{n}\expval{A_{n}^{2}}_{\rho}\,.
    \label{eq:anticomm theorem}
\end{align}
A simple example for the applicability of this theorem is the set of single-qubit Pauli operators $\{X,Y,Z\}$. Since all of these operators anticommute and square to the identity, the ACT then simply requires that
\begin{align}
    \expval{X}^{2}_{\rho}\,+\,\expval{Y}^{2}_{\rho}\,+\,\expval{Z}^{2}_{\rho} &\leq\,\expval{\mathds{1}}_{\rho}\,=\,1.
    \label{eq:ACT for single qubit Paulis}
\end{align}
In other words, for single-qubit Pauli operators, the ACT is equivalent to demanding that Bloch vectors are (sub)normalized, i.e., positivity of the density operator~$\rho$. A less trivial example of the ACT arises for 2 qubits. Consider the set of operators
\begin{align}
    \{X_{1}X_{2},\,Y_{1}Y_{2},\,Z_{1}Z_{2},\,X_{2}X_{3},\,Y_{2}Y_{3},\,Z_{2}Z_{3}\}\,,
    \label{eq:nearest neighbour Bell fid operators 3 qubits}
\end{align}
where the shorthand notation for $N$-qubit operators is
\begin{align}
    O_{i}   &\equiv \mathds{1}_{1}\otimes\ldots\otimes\mathds{1}_{i-1}\otimes O_{i}\otimes\mathds{1}_{i+1}\otimes\ldots\otimes\mathds{1}_{N},
    \label{eq:N qubit operator shorthand}
\end{align}
and $O\in\{X,Y,Z\}$. We can sort the six operators in the set displayed in Eq.~(\ref{eq:nearest neighbour Bell fid operators 3 qubits}) into three pairs of anticommuting operators, e.g.,
\begin{subequations}
\label{eq:anticommuting bipartite Paulis sets}
\begin{align}
    \{X_{1}X_{2},Y_{2}Y_{3}\}_{+}    &=0\,,\\
    \{Y_{1}Y_{2},Z_{2}Z_{3}\}_{+}    &=0\,,\\
    \{Z_{1}Z_{2},X_{2}X_{3}\}_{+}    &=0.
\end{align}
\end{subequations}
Since the spectra of all six operators are $\{\pm1\}$ (with twofold degeneracy), we further have $\expval{O_{i}O_{i+1}}^{2}\leq1$, and the ACT theorem hence tells us that
\begin{subequations}
\label{eq:anticommuting bipartite Paulis sets ACT}
\begin{align}
    \expval{X_{1}X_{2}}_{\rho}^{2}\,+\,\expval{Y_{2}Y_{3}}_{\rho}^{2}    &\leq1\,,\\
    \expval{Y_{1}Y_{2}}_{\rho}^{2}\,+\,\expval{Z_{2}Z_{3}}_{\rho}^{2}    &\leq1\,,\\
    \expval{Z_{1}Z_{2}}_{\rho}^{2}\,+\,\expval{X_{2}X_{3}}_{\rho}^{2}   &\leq1\,.
\end{align}
\end{subequations}
To see where these bounds can be of use, let us next examine fidelities with $2$-qubit Bell states.


\subsubsection{Average Bell State Fidelities}\label{sec:average bell state fidelities}

We now want to consider ways of quantifying how close a given $2$-qubit state is to a maximally entangled Bell state. To this end, note that the density operators for the four Bell states can be written in a generalized Bloch decomposition as
\begin{subequations}
\label{eq:Bell state density operator Bloch decomp}
\begin{align}
    \rho_{\psi^{-}} &=\ket{\psi^{-}}\!\!\bra{\psi^{-}}=\tfrac{1}{4}\bigl(\mathds{1}_{1,2}-X_{1}X_{2}-Y_{1}Y_{2}-Z_{1}Z_{2}\bigr),
    \label{eq:psi minus bloch decomp}\\
    \rho_{\psi^{+}} &=\ket{\psi^{+}}\!\!\bra{\psi^{+}}=\tfrac{1}{4}\bigl(\mathds{1}_{1,2}+X_{1}X_{2}+Y_{1}Y_{2}-Z_{1}Z_{2}\bigr),
    \label{eq:psi plus bloch decomp}\\
    \rho_{\phi^{-}} &=\ket{\phi^{-}}\!\!\bra{\phi^{-}}=\tfrac{1}{4}\bigl(\mathds{1}_{1,2}-X_{1}X_{2}+Y_{1}Y_{2}+Z_{1}Z_{2}\bigr),
    \label{eq:phi minus bloch decomp}\\
    \rho_{\phi^{+}} &=\ket{\phi^{+}}\!\!\bra{\phi^{+}}=\tfrac{1}{4}\bigl(\mathds{1}_{1,2}+X_{1}X_{2}-Y_{1}Y_{2}+Z_{1}Z_{2}\bigr).
    \label{eq:phi plus bloch decomp}
\end{align}
\end{subequations}
For any $2$-qubit density operator $\rho$, we can then compute the fidelity with any of the Bell states. For this purpose we use the Uhlmann fidelity $\mathcal{F}$, given by
\begin{align}
    \mathcal{F}(\rho,\sigma)    &=\,\Bigl(\tr\sqrt{\sqrt{\sigma}\rho\sqrt{\sigma}}\Bigr)^{2}\,,
    \label{eq:Uhlmann fidelity general}
\end{align}
which reduces to
\begin{align}
    \mathcal{F}(\rho,\ket{\psi}\!\!\bra{\psi})  &=\,\bra{\psi}\rho\ket{\psi}=\tr\bigl(\rho\ket{\psi}\!\!\bra{\psi}\bigr)
    \label{eq:Uhlmann fidelity pure state}
\end{align}
if one of the arguments is a pure state. For example, for the Bell state $\ket{\psi^{-}}$ one can use Eq.~(\ref{eq:psi minus bloch decomp}) and the fact that all Pauli operators are traceless to find
\begin{align}
    \mathcal{F}(\rho,\rho_{\psi^{-}})  &=\,
    \tfrac{1}{4}\bigl(1-\expval{X_{1}X_{2}}_{\rho}-\expval{Y_{1}Y_{2}}_{\rho}-\expval{Z_{1}Z_{2}}_{\rho}\bigr).
    \label{eq:Uhlmann fidelity psi minus}
\end{align}
Since the only difference to the fidelities with any of the other Bell states are the relative signs between the different expectation values, we can immediately note that the fidelity of $\rho$ with any of the four Bell states is bounded according to
\begin{align}
    \mathcal{F}(\rho,\rho_{\mathrm{Bell}})  &\leq\,
    \tfrac{1}{4}\bigl(1+|\!\expval{X_{1}X_{2}}\!|+|\!\expval{Y_{1}Y_{2}}\!|+|\!\expval{Z_{1}Z_{2}}\!|\bigr),
    \label{eq:Uhlmann fidelity with arbitrary Bell state}
\end{align}
where we have dropped the subscript for the state $\rho$ on the expectation values for brevity.


\subsubsection{Nearest-Neighbour Average Bell Fidelity}

When the system consists of more than 2 qubits, we can evaluate the fidelity with $2$-qubit Bell states for any two of the constituent qubits. For simplicity, let us first consider the nearest neighbours for now and examine the case of 3 qubits. The average fidelity with arbitrary nearest-neighbour Bell states is then
\begin{align}
    & \tfrac{1}{2}\bigl(\mathcal{F}(\rho,\rho_{\mathrm{Bell},12})+\mathcal{F}(\rho,\rho_{\mathrm{Bell},23})\bigr)    \,\leq\,\bar{\mathcal{F}}_{\mathrm{NN~Bell}},
    \label{eq:3QB average Bell fid bound}
\end{align}
where we define the quantity $\bar{\mathcal{F}}_{\mathrm{NN~Bell}}$ as the upper bound
\begin{align}
    \bar{\mathcal{F}}_{\mathrm{NN~Bell}}&:=\tfrac{1}{8}\bigl(2+|\!\expval{X_{1}X_{2}}\!|+|\!\expval{Y_{1}Y_{2}}\!|+|\!\expval{Z_{1}Z_{2}}\!|
    \nonumber\\[1mm]
    &\ \ \ \ +|\!\expval{X_{2}X_{3}}\!|+|\!\expval{Y_{2}Y_{3}}\!|+|\!\expval{Z_{2}Z_{3}}\!|\bigr),
    \label{eq:3QB average Bell fid}
\end{align}
but we refer to $\bar{\mathcal{F}}_{\mathrm{NN~Bell}}$ as the average nearest neighbour Bell fidelity from now on for simplicity. Next, we make use of the relation between the $1$-norm $|\hspace*{-1pt}|\vec{a}|\hspace*{-1pt}|_{1}=\sum_{i=1}^{n}|a_{i}|$ and the $2$-norm $|\hspace*{-1pt}|\vec{a}|\hspace*{-1pt}|_{2}=\bigl(\sum_{i=1}^{n}|a_{i}|^{2}\bigr)^{1/2}$ in an $n$-dimensional vector space, i.e., the fact that
\begin{align}
    |\hspace*{-1pt}|\vec{a}|\hspace*{-1pt}|_{1} &=\sum_{i=1}^{n}|a_{i}|\times1=|(\vec{a},\vec{1})|
    \leq
    |\hspace*{-1pt}|\vec{a}|\hspace*{-1pt}|_{2}\bigl(\sum_{i=1}^{n}1^{2}\bigr)^{1/2}=\sqrt{n}|\hspace*{-1pt}|\vec{a}|\hspace*{-1pt}|_{2},
    \label{eq:relation between 1norm and 2norm}
\end{align}
where we have taken $\vec{1}=(1,1,\ldots,1)^{T}$ to be a vector whose components (w.r.t. whichever basis is chosen for $\vec{a}$) are all equal to $1$, and we have used the Cauchy-Schwarz inequality $|(\vec{a},\vec{b})|\leq |\hspace*{-1pt}|\vec{a}|\hspace*{-1pt}|_{2} |\hspace*{-1pt}|\vec{b}|\hspace*{-1pt}|_{2}$. Combining this with the ACT theorem from Eq.~(\ref{eq:anticomm theorem}) we find, e.g.,
\begin{align}
    |\!\expval{X_{1}X_{2}}\!|+|\!\expval{Y_{2}Y_{3}}\!|    &\leq
    \sqrt{2}\bigl(\expval{X_{1}X_{2}}^{2}\,+\,\expval{Y_{2}Y_{3}}^{2}\bigr)\leq\sqrt{2}.
    \label{eq:bound trick}
\end{align}
Applying the same procedure to the other pairs of expectation values of anticommuting operators in Eq.~(\ref{eq:3QB average Bell fid}), we arrive at the bound
\begin{align}
    \bar{\mathcal{F}}_{\mathrm{NN~Bell}}    &\leq\tfrac{1}{8}\bigl(2+3\sqrt{2}\bigr).
    \label{eq:3QB average Bell fid AC bound}
\end{align}

The average nearest-neighbour Bell state fidelity can of course be generalized to $N$ qubits, i.e., the upper bound on the average nearest-neighbour Bell fidelity is
\begin{align}
    &\frac{1}{N-1}\sum\limits_{i=1}^{N-1}\mathcal{F}(\rho,\rho_{\mathrm{Bell},i(i+1)})  \leq
    \bar{\mathcal{F}}^{(N)}_{\mathrm{NN~Bell}},
\end{align}
where we have defined
\begin{align}
    &\bar{\mathcal{F}}^{(N)}_{\mathrm{NN~Bell}}\,:=\,\tfrac{1}{4(N-1)}\Bigl((N-1)+\sum\limits_{i=1}^{N-1}\sum\limits_{O=X,Y,Z}|\!\expval{O_{1}O_{i+1}}\!|\Bigr)
    \nonumber\\[1mm]
    &\ =\tfrac{1}{4(N-1)}\Bigl((N-1)+
    |\!\expval{X_{1}X_{2}}\!|+|\!\expval{Y_{1}Y_{2}}\!|+|\!\expval{Z_{1}Z_{2}}\!|
    \nonumber\\[1mm]
    &\ \ \ \ \ +|\!\expval{X_{2}X_{3}}\!|+|\!\expval{Y_{2}Y_{3}}\!|+|\!\expval{Z_{2}Z_{3}}\!|+\dots\nonumber\\[1mm]
    &\ \ \ \dots+|\!\expval{X_{N-1}X_{N}}\!|+|\!\expval{Y_{N-1}Y_{N}}\!|+|\!\expval{Z_{N-1}Z_{N}}\!|\Bigr).
    \label{eq:NQB average Bell fid}
\end{align}
The expression on the right-hand side contains $N-1$ triples of expectation values. If $N$ is odd, then $N-1$ is even, and each expectation value of an operator $O_{i}O_{i+1}$ can be paired with another expectation value of an operator $O\pr_{i+1}O\pr_{i+2}$ that anticommutes with it, i.e., $O,O\pr\in\{X,Y,Z\}$ and $\{O_{i}O_{i+1},O\pr_{i+1}O\pr_{i+2}\}_{+}=0$. The bound of Eq.~(\ref{eq:bound trick}) can hence be used $\tfrac{3(N-1)}{2}$ times, and we arrive at
\begin{align}
    \bar{\mathcal{F}}^{(N \mathrm{odd})}_{\mathrm{NN~Bell}}    &\leq\tfrac{1}{4(N-1)}\Bigl((N-1)+\tfrac{3(N-1)}{2}\sqrt{2}\Bigr)=\tfrac{1}{8}\bigl(2+3\sqrt{2}\bigr).
\end{align}
However, when $N$ is even, one triple of expectation values (w.l.o.g. for $i=N-1$) remains unpaired and can only be bounded by
\begin{align}
    |\!\expval{X_{N-1}X_{N}}\!|+|\!\expval{Y_{N-1}Y_{N}}\!|+|\!\expval{Z_{N-1}Z_{N}}\!| &\leq3\,.
\end{align}
Thus we arrive at the following upper bound on the nearest-neighbour average Bell fidelity for arbitrary $N$-qubit states, i.e.,
\begin{align}
    \bar{\mathcal{F}}^{(N)}_{\mathrm{NN~Bell}}    &\leq
    \begin{cases}
        \tfrac{1}{8}\bigl(2+3\sqrt{2}\bigr) &   \mbox{($N$ odd)}\\
        \tfrac{1}{4(N-1)}\Bigl(\!(N\!-\!1)+\tfrac{3(N-2)\sqrt{2}}{2}+3\Bigr) &   \mbox{($N$ even)}
    \end{cases}.
    \label{eq:NQB nearest neighbour Bell bound general}
\end{align}


\subsubsection{Symmetric Average Bell Fidelity}

Instead of restricting the analysis to nearest neighbours as in Eq.~(\ref{eq:3QB average Bell fid}), one can of course also average over all pairings of 2 qubits, obtaining a quantity that is symmetric w.r.t. the exchange of any 2 qubits. Noting that there are $b_{N}=\tbinom{N}{2}=\tfrac{1}{2}\tfrac{N!}{(N-2)!}$ different such pairings, we have the upper bound
\begin{align}
    \tfrac{1}{b_{N}}\sum\limits_{\substack{i,j=1 \\ i<j}}^{N}\mathcal{F}(\rho,\rho_{\mathrm{Bell},ij})
    &\leq\bar{\mathcal{F}}^{(N)}_{\mathrm{Bell}},
    \label{eq:NQB symm average Bell fid bound}
\end{align}
with the definition
\begin{align}
    \bar{\mathcal{F}}^{(N)}_{\mathrm{Bell}}& :=
    \tfrac{1}{4b_{N}}\Bigl(b_{N}+\!\!\sum\limits_{\substack{i,j=1 \\ i<j}}^{N}\sum\limits_{O=X,Y,Z}|\!\expval{O_{i}O_{j}}\!|\Bigr).
    \label{eq:NQB symm average Bell fid}
\end{align}
For, instance, for 3 qubits we have $b_{3}=3$ and the symmetric average Bell fidelity reads
\begin{align}
    \bar{\mathcal{F}}^{(3)}_{\mathrm{Bell}} &\ =\tfrac{1}{12}\Bigl(3
    +|\!\expval{X_{1}X_{2}}\!|+|\!\expval{X_{2}X_{3}}\!|+|\!\expval{X_{1}X_{3}}\!|\nonumber\\[1mm]
    &\hspace*{1.35cm}
    +\,|\!\expval{Y_{1}Y_{3}}\!|\ +\ |\!\expval{Y_{1}Y_{2}}\!|\,+\,|\!\expval{Y_{2}Y_{3}}\!|\nonumber\\[1mm]
    &\hspace*{1.35cm}
    +|\!\expval{Z_{2}Z_{3}}\!|\,+\,|\!\expval{Z_{1}Z_{3}}\!|\,+\,|\!\expval{Z_{1}Z_{2}}\!|\Bigr).
    \label{eq:3QB symm average Bell fid}
\end{align}
Here, we have arranged the expectation values such that it becomes immediately obvious that the triples of operators corresponding to expectation values listed directly below or above each other mutually anticommute. We can then apply the bound of Eq.~(\ref{eq:relation between 1norm and 2norm}) and the ACT of Eq.~(\ref{eq:anticomm theorem}), e.g., as illustrated for the terms
\begin{align}
    & |\!\expval{X_{1}X_{2}}\!|+|\!\expval{Y_{1}Y_{3}}\!|+|\!\expval{Z_{2}Z_{3}}\!|   \label{eq:trick for three qubits}\\[1mm]
    &\ \ \leq\sqrt{3}\Bigl(|\!\expval{X_{1}X_{2}}\!|^{2}+|\!\expval{Y_{1}Y_{3}}\!|^{2}+|\!\expval{Z_{2}Z_{3}}\!|^{2}\Bigr)\leq\sqrt{3}.
    \nonumber
\end{align}
We thus arrive at the bound
\begin{align}
    \bar{\mathcal{F}}^{(3)}_{\mathrm{Bell}} &\ \leq\tfrac{1}{12}\bigl(3+3\sqrt{3}\bigr)=\tfrac{1}{4}\bigl(1+\sqrt{3}\bigr).
    \label{eq:3QB symm average Bell fid bound}
\end{align}
In fact, the same bound applies for arbitrary numbers of qubits, since all $3b_{N}$ expectation values can be collected in groups of 3 mutually anticommuting operators. To see this, we use an inductive proof. Assume that we have found $b_{N}$ groups of three anticommuting operators for $N\geq3$ qubits and we wish to add another qubit. This means that we have to additionally consider the operators
\begin{center}
\begin{tabular}{c c c c c}
    $X_{1}X_{N+1}$,    &   $X_{2}X_{N+1}$,    &   $X_{3}X_{N+1}$,    &   $\ldots$ ,  &   $X_{N}X_{N+1}$,\\[1mm]
    $Y_{N}Y_{N+1}$,    &   $Y_{1}Y_{N+1}$,    &   $Y_{2}Y_{N+1}$,    &   $\ldots$ ,  &   $Y_{N-1}Y_{N+1}$,\\[1mm]
    $Z_{N-1}Z_{N+1}$,    &   $Z_{N}Z_{N+1}$,    &   $Z_{1}Z_{N+1}$,    &   $\ldots$ ,  &   $Z_{N-2}Z_{N+1}$.
\end{tabular}
\end{center}
All columns contain three mutually anticommuting operators for $N\geq3$. If the original $3b_{N}$ operators can be arranged in mutually anticommuting triples, then also the new set of $3b_{N+1}$ operators can be grouped in this way, which concludes the inductive step. We have already demonstrated that this statement is true for $N=3$ and have hence shown that for any $N\geq3$ we have the bound
\begin{align}
    \bar{\mathcal{F}}^{(N\geq3)}_{\mathrm{Bell}} &\ \leq\bar{\mathcal{F}}^{(N) \mathrm{max}}_{\mathrm{Bell}}:=\tfrac{1}{4}\bigl(1+\sqrt{3}\bigr).
    \label{eq:NQB symm average Bell fid bound}
\end{align}

Having established these general bounds that apply for arbitrary quantum states, we next examine how these bounds can be improved upon when the states in question are biseparable. This will allow us to formulate criteria for the detection of genuine multipartite entanglement.


\subsection{GME Witnesses}\label{sec:GME witnesses}

In this section, we establish upper bounds for the nearest-neighbour and symmetric average Bell fidelities for biseparable states. These new upper bounds are below the respective bounds of Eqs.~(\ref{eq:NQB nearest neighbour Bell bound general}) and~(\ref{eq:NQB symm average Bell fid bound}) and hence leave room for GME states in between. That is, any states for which the combinations of expectation values discussed above provide values beyond these biseparability bounds are GME. As we shall see, the biseparability bounds for nearest-neighbour Bell fidelities are not directly useful for detecting GME, but these bounds serve as a simple example for discussing the method of construction which will be helpful for identifying GME witnesses based on symmetric average Bell fidelities.


\subsubsection{Outline of the Technique}

In the following we consider bipartitions $A|B$ of the set $\kappa=\{1,2,\ldots,N\}$ of all $N$ qubits, that is, we split $\kappa$ into two sets,
\begin{subequations}
\begin{align}
    A   &=\{a_{1},a_{2},\ldots,a_{k}|a_{i}\in\kappa, a_{i}\neq a_{j}\forall i\neq j\},\\[1mm]
    B   &=\{b_{1},b_{2},\ldots,b_{N-k}|b_{i}\in\kappa, b_{i}\neq b_{j}\forall i\neq j\},
\end{align}
\end{subequations}
such that $A\cup B=\kappa$ and $A\cap B=\emptyset$. For $N$ qubits, one has $2^{N-1}-1$ different bipartitions.

Before we continue, let us briefly recall the definitions of biseparability and genuine multipartite entanglement. In general, a pure, $N$-partite state $\ket{\psi}\in\mathcal{H}\subtiny{0}{0}{1,2,\ldots,N}=\mathcal{H}\subtiny{0}{0}{1}\otimes\mathcal{H}\subtiny{0}{0}{2}\otimes\ldots\mathcal{H}\subtiny{0}{0}{N}$ is called $k$-separable if it can be written as a tensor product with respect to some partition of $\mathcal{H}\subtiny{0}{0}{1,2,\ldots,N}$ into~$k\leq N$ subsystems. As a special case of this definition, a pure state is called biseparable, if it can be written as a tensor product w.r.t. some bipartition, i.e., if there exists a bipartition $A|B$ such that $\ket{\psi}=\ket{\phi}_{A}\ket{\chi}_{B}$. Conversely, a pure state $\ket{\psi}\in\mathcal{H}\subtiny{0}{0}{1,2,\ldots,N}$ that is not biseparable is called \emph{genuinely $N$-partite entangled}. A mixed state with density operator $\rho$ is considered to be genuinely multipartite entangled if it cannot be written as a convex combination of biseparable states, that is, if it cannot be written as
\begin{align}
    \rho_{\mathrm{bisep}}   &=\,\sum\limits_{i}p_{i}\ket{\psi_{\mathrm{bisep}}\suptiny{0}{0}{(i)}}\!\!\bra{\psi_{\mathrm{bisep}}\suptiny{0}{0}{(i)}},
\end{align}
where $\sum_{i}p_{i}=1$ with $0\leq p_{i}\leq1$ and $\ket{\psi_{\mathrm{bisep}}\suptiny{0}{0}{(i)}}$ are biseparable pure states. Note that the $\ket{\psi_{\mathrm{bisep}}\suptiny{0}{0}{(i)}}$ for different $i$ can be separable w.r.t. different bipartitions.

Now, consider a bipartition $A|B$ and an operator $O_{i}O_{j}$ such that $i\in A$ and $j\in B$. If the system is in a pure state $\ket{\psi}$ that is separable w.r.t. to this bipartition, i.e., if $\ket{\psi}_{AB}=\ket{\phi}_{A}\ket{\chi}_{B}$, then we have
\begin{align}
    \expval{O_{i}O_{j}}_{\psi}  &=\,\expval{O_{i}}_{\phi}\expval{O_{j}}_{\chi}\,.
\end{align}
When we have a triple of operators $X_{i}X_{j}$, $Y_{i}Y_{j}$, and $Z_{i}Z_{j}$ for such a separable state across $A|B$, we have
\begin{align}
    & |\!\expval{X_{i}X_{j}}\!|+|\!\expval{Y_{i}Y_{j}}\!|+|\!\expval{Z_{i}Z_{j}}\!|
    \label{eq:bisep trick}\\[1mm]
    &\ =\,|\!\expval{X_{i}}\!|.|\!\expval{X_{j}}\!|+|\!\expval{Y_{i}}\!|.|\!\expval{Y_{j}}\!|+|\!\expval{Z_{i}}\!|.|\!\expval{Z_{j}}\!|
    \nonumber\\[1mm]
    &\ \leq\,\prod_{n=i,j}\sqrt{|\!\expval{X_{n}}\!|^{2}+|\!\expval{Y_{n}}\!|^{2}+|\!\expval{Z_{n}}\!|^{2}}\,\leq\,1,\nonumber
\end{align}
where we have used the Cauchy-Schwarz inequality in the second-to-last step and the subnormalization of the Bloch vector in the last step. The inequality~(\ref{eq:bisep trick}) can be used to bound the Bell fidelities for pure biseparable states for different bipartitions.

As an example, consider again the nearest-neighbour average Bell fidelity for 3 qubits from Eq.~(\ref{eq:3QB average Bell fid}). For the bipartition $1|23$, we can apply~(\ref{eq:bisep trick}) to the first three expectation values in Eq.~(\ref{eq:3QB average Bell fid}), while the remaining three can each be bounded by $1$. A similar argument can be made for the bipartition $12|3$ by exchanging the roles of the two triples of expectation values, such that
\begin{align}
    \bar{\mathcal{F}}_{\mathrm{NN~Bell}}^{1|23, 12|3}   &\leq\tfrac{1}{8}\bigl(2+1+3\bigr)=\tfrac{3}{4},
    \label{eq:3QB average Bell fid bipartitions 1vs23 and 12vs3}
\end{align}
where the superscripts indicate that the inequality is satisfied for states that are biseparable w.r.t. (at least one of) the listed bipartitions.
When we examine the bipartition $2|13$, the situation is slightly different, since Eq.~(\ref{eq:bisep trick}) can be used for all expectation values and we have
\begin{align}
    \bar{\mathcal{F}}_{\mathrm{NN~Bell}}^{2|13}   &\leq\tfrac{1}{8}\bigl(2+1+1\bigr)=\tfrac{1}{2}.
    \label{eq:3QB average Bell fid bipartitions 2vs13}
\end{align}
Any pure 3-qubit state that is separable w.r.t. one or more of these bipartitions (any pure, biseparable state of 3 qubits) must hence satisfy $\bar{\mathcal{F}}_{\mathrm{NN~Bell}}\leq\tfrac{3}{4}$. Moreover, since any mixed state is considered to be biseparable when it can be written as a convex combination of biseparable pure states (not necessarily w.r.t. to the same bipartition), all mixed, biseparable states must also respect this bound. Conversely, the first 3 qubits of any state $\rho$ for which
\begin{align}
    &\tfrac{1}{8}\bigl(2+|\!\expval{X_{1}X_{2}}\!|+|\!\expval{Y_{1}Y_{2}}\!|+|\!\expval{Z_{1}Z_{2}}\!|
    \nonumber\\[1mm]
    &\ \ \ \ +|\!\expval{X_{2}X_{3}}\!|+|\!\expval{Y_{2}Y_{3}}\!|+|\!\expval{Z_{2}Z_{3}}\!|\bigr)\,>\,\tfrac{3}{4}
    \label{eq:3QB GME witness nearest neighbour}
\end{align}
are genuinely 3-partite entangled.


\subsubsection{Nearest-Neighbour Average Bell Fidelity as GME Witness}\label{sec:Nearest-Neighbour Average Bell Fidelity as GME Witness}

In principle, the nearest-neighbour Bell fidelity could hence provide a detection criterion for GME that can be generalized to $N$ qubits. However, at this point a remark on the detection power of this quantity is in order, since even some paradigmatic cases of genuinely tripartite entangled states for 3 qubits cannot be detected with this bound. That is, for the 3-qubit GHZ and W states $\ket{\psi_{\mathrm{GHZ}}\suptiny{0}{0}{(3)}}$ and $\ket{\psi_{\mathrm{W}}\suptiny{0}{0}{(3)}}$ (or the local unitarily equivalent $2$-excitation Dicke state $\ket{\psi_{\mathrm{D},2}\suptiny{0}{0}{(3)}}$), given by
\begin{subequations}
\label{eq:three QB GME states}
\begin{align}
    \ket{\psi_{\mathrm{GHZ}}\suptiny{0}{0}{(3)}}    &=\,\tfrac{1}{\sqrt{2}}\bigl(\ket{000}+\ket{111}\bigr),\\[1mm]
    \ket{\psi_{\mathrm{W}}\suptiny{0}{0}{(3)}}      &=\,\tfrac{1}{\sqrt{3}}\bigl(\ket{100}+\ket{010}+\ket{001}\bigr),\\[1mm]
    \ket{\psi_{\mathrm{D},2}\suptiny{0}{0}{(3)}}    &=\,\tfrac{1}{\sqrt{3}}\bigl(\ket{110}+\ket{101}+\ket{011}\bigr),
\end{align}
\end{subequations}
one finds nearest-neighbour average Bell fidelities of
\begin{subequations}
\label{eq:three QB GME states NN average Bell fidelities}
\begin{align}
\bar{\mathcal{F}}_{\mathrm{NN~Bell}}^{(3)}(\ket{\psi_{\mathrm{GHZ}}\suptiny{0}{0}{(3)}}) &=\tfrac{1}{2},\\[1mm]
\bar{\mathcal{F}}_{\mathrm{NN~Bell}}^{(3)}(\ket{\psi_{\mathrm{W}}\suptiny{0}{0}{(3)}})   &=\tfrac{2}{3},\\[1mm]
\bar{\mathcal{F}}_{\mathrm{NN~Bell}}^{(3)}(\ket{\psi_{\mathrm{D},2}\suptiny{0}{0}{(3)}}) &=\tfrac{2}{3},
\end{align}
\end{subequations}
whereas the corresponding bound for detecting GME is $\tfrac{3}{4}$.

One can hence try to improve the method or find an alternative. One way to improve the bound is by way of taking into account the purity of the biseparable states. That is, if we consider again the worst-case bipartition $1|23$ for 3 qubits under the assumption that the state is separable w.r.t. this cut, i.e., that $\ket{\psi}_{123}=\ket{\phi}_{1}\ket{\chi}_{23}$, we have
\begin{align}
    \bar{\mathcal{F}}_{\mathrm{NN~Bell}}^{1|23}   &\leq\tfrac{1}{8}\bigl(2+
    \sqrt{|\!\expval{X_{2}}\!|^{2}+|\!\expval{Y_{2}}\!|^{2}+|\!\expval{Z_{2}}\!|^{2}}
    \label{eq:3QB average Bell fid bipartitions 1vs23 again}\\[1mm]
    &+|\!\expval{X_{2}X_{3}}\!|+|\!\expval{Y_{2}Y_{3}}\!|+|\!\expval{Z_{2}Z_{3}}\!|\bigr)
    \nonumber\\[1mm]
    &=\tfrac{1}{8}\bigl(2+|\vec{b}|+\sum\limits_{n=1}^{3}|t_{nn}|\bigr),\nonumber
\end{align}
where we have used the Bloch vector $\vec{b}$ of the second qubit and the correlation tensor $t=(t_{mn})$ of qubits $2$ and $3$. In other words, the state $\ket{\chi}_{23}$ can be written in a generalized Bloch decomposition as
\begin{align}
    \rho_{\chi}=\ket{\chi}\!\!\bra{\chi}  &=\tfrac{1}{4}\Bigl(\mathds{1}+\vec{b}\cdot\vec{\sigma}\otimes\mathds{1}+\mathds{1}\otimes\vec{c}\cdot\vec{\sigma}
    +\!\!\!\!\sum\limits_{i,j=1}^{3}\!\!t_{ij}\sigma_{i}\otimes\sigma_{j}\Bigr),
\end{align}
where $\vec{\sigma}=(\sigma_{n})$ is the vector of Pauli operators $(\sigma_{1}=X,\sigma_{2}=Y,\sigma_{3}=Z)$. Now, since $\ket{\chi}_{23}$ is a pure state, we have $\tr(\rho_{\chi}^{2})=1$, which translates to
\begin{align}
    \tfrac{1}{4}\Bigl(1+|\vec{b}|^{2}+|\vec{c}|^{2}+\sum\limits_{m,n=1}^{3}|t_{mn}|^{2}\Bigr)   &=1,
\end{align}
and we can hence derive the bound
\begin{align}
    |\vec{b}|^{2}+\sum\limits_{n=1}^{3}|t_{nn}|^{2}   &\leq 3.
    \label{eq:trick for improved bound}
\end{align}
Interpreting $|\vec{b}|$ and $|t_{nn}|$ ($n=1,2,3$) as coordinates in $\mathbb{R}^{4}$, we find that Eq.~(\ref{eq:trick for improved bound}) defines a four-dimensional sphere of radius $\sqrt{3}$. The sum of the coordinates is then maximal when all coordinates take the same value $\sqrt{3/4}$. Inserting into Eq.~(\ref{eq:3QB average Bell fid bipartitions 1vs23 again}), we then get the bound
\begin{align}
    \bar{\mathcal{F}}_{\mathrm{NN~Bell}}^{1|23}   &\leq\tfrac{1}{8}\bigl(2+4\sqrt{\tfrac{3}{4}}\bigr)=\tfrac{1}{4}\bigl(1+\sqrt{3}\bigr)\approx0.683013.
    \label{eq:3QB average Bell fid bipartitions 1vs23 improved bound}
\end{align}
Since the bipartition $12|3$ is equivalent and for $2|13$ we have the lower value $\bar{\mathcal{F}}_{\mathrm{NN~Bell}}^{2|13}\leq\tfrac{1}{2}$, the ``improved" biseparability bound for the nearest-neighbour average Bell fidelity for three qubits is $\tfrac{1}{4}\bigl(1+\sqrt{3}\bigr)\approx0.683013$. This value is still above the fidelity $\tfrac{2}{3}$ obtained for pure GME states of 3 qubits in Eq.~(\ref{eq:three QB GME states NN average Bell fidelities}). In addition, we have also conducted a numerical search which did not reveal any pure 3-qubit states with nearest-neighbour average Bell fidelities beyond $\tfrac{2}{3}$. At the same time, one can find biseparable states that give values for $\bar{\mathcal{F}}_{\mathrm{NN~Bell}}^{(3)}$ that are very close to $\tfrac{2}{3}$. For instance, for the state $\rho_{\mathrm{bisep}}=\ket{0}\!\!\bra{0}_{1}\otimes\tilde{\rho}_{23}$, where $\ket{0}_{1}$ is an eigenstate of $Z$, and the (nearly pure) state $\tilde{\rho}$ has Bloch vectors $\vec{b}=\vec{c}=(0,0,0.447)^{T}$ and a diagonal correlation matrix $t=\diag\{0.894,-0.894,1\}$, we find $\bar{\mathcal{F}}_{\mathrm{NN~Bell}}^{(3)}=0.654375$. We therefore conclude that, in their present form, GME witnesses based on the nearest-neighbour average Bell fidelity are practically irrelevant for 3 qubits and there is no reason to expect an improvement for more than 3 qubits.

We therefore now continue with an analysis of a different quantity, the symmetric average Bell fidelity.


\subsubsection{Symmetric Average Bell Fidelity as GME Witness}\label{sec:Symmetric Average Bell Fidelity as GME Witness}

In this section, we discuss the usefulness of the symmetric average Bell fidelity as a witness for GME. To this end, we again need to identify the bipartitions providing the worst (largest) upper bound for $\bar{\mathcal{F}}^{(N)}_{\mathrm{Bell}}$ under the assumption of separability w.r.t. to the respective bipartition. Since the combination of expectation values that we consider now is symmetric under the exchange of any 2 qubits, this task is rather straightforward.

First, we consider the case of 3 qubits separately, where all three possible bipartitions (i.e., $1|23$, $2|13$, and $12|3$) are equivalent. If the system state is pure and separable w.r.t. to any of these bipartitions, two of the triples of expectation values in Eq.~(\ref{eq:3QB symm average Bell fid}) are ``cut" by the bipartition and can be bounded by $1$, while the remaining triple consists of three commuting observables, whose expectation values are jointly bounded by $3$. For any labelling of the 3 qubits we hence have
\begin{align}
    \bar{\mathcal{F}}^{1|23}_{\mathrm{Bell}}    &\leq\,\tfrac{1}{12}\bigl(3+1+1+3\bigr)=\tfrac{2}{3}.
\end{align}
As we discussed in Section~\hyperref[sec:Nearest-Neighbour Average Bell Fidelity as GME Witness]{A.II.2}, this bound has to be compared with values achievable with pure GME states. For the 3-qubit GHZ-, W-, and $2$-excitation Dicke states, we find symmetric average Bell fidelities
\begin{subequations}
\label{eq:three QB GME states NN average Bell fidelities}
\begin{align}
\bar{\mathcal{F}}_{\mathrm{Bell}}^{(3)}(\ket{\psi_{\mathrm{GHZ}}\suptiny{0}{0}{(3)}}) &=\tfrac{1}{2},\\[1mm]
\bar{\mathcal{F}}_{\mathrm{Bell}}^{(3)}(\ket{\psi_{\mathrm{W}}\suptiny{0}{0}{(3)}})   &=\tfrac{2}{3},\\[1mm]
\bar{\mathcal{F}}_{\mathrm{Bell}}^{(3)}(\ket{\psi_{\mathrm{D},2}\suptiny{0}{0}{(3)}}) &=\tfrac{2}{3},
\end{align}
\end{subequations}
which happen to coincide with the corresponding nearest-neighbour average Bell fidelities of Eq.~(\ref{eq:three QB GME states NN average Bell fidelities}). We must hence try to improve the bound using a similar trick as before in Appendix~\hyperref[sec:Nearest-Neighbour Average Bell Fidelity as GME Witness]{A.II.2}. Again assuming a biseparable pure state for the bipartition $1|23$, we can write
\begin{align}
    \bar{\mathcal{F}}_{\mathrm{Bell}}^{1|23}   &\leq\tfrac{1}{12}
    \bigl(3+|\vec{b}|+|\vec{c}|+\sum\limits_{n=1}^{3}|t_{nn}|\bigr)
    \label{eq:3QB average symm Bell fid bipartitions 1vs23 again}
\end{align}
and in analogy to Eq.~(\ref{eq:trick for improved bound}) we can bound each of the moduli $|\vec{b}|$, $|\vec{c}|$, and $|t_{nn}|$ (for $n=1,2,3$) by $\sqrt{3/5}$, which gives the bound
\begin{align}
    \bar{\mathcal{F}}_{\mathrm{Bell}}^{(3) \mathrm{bisep}}   &\leq\tfrac{1}{12}
    \bigl(3+5\sqrt{\tfrac{3}{5}}\bigr)=\tfrac{1}{12}\bigl(3+\sqrt{15}\bigr)\approx0.572749.
    \label{eq:3QB average symm Bell fid bound improved}
\end{align}
Using numerical optimisation, we can also provide a pure biseparable state that comes very close to this bound. That is, for the state $\rho_{\mathrm{bisep}}=\ket{0}\!\!\bra{0}_{1}\otimes\tilde{\rho}_{23}$, where $\ket{0}_{1}$ is an eigenstate of $Z$, and the pure state $\tilde{\rho}$ has Bloch vectors $\vec{b}=\vec{c}=(0,0,\tfrac{1}{\sqrt{2}})^{T}$ and a diagonal correlation matrix $t=\diag\{\tfrac{1}{\sqrt{2}},-\tfrac{1}{\sqrt{2}},1\}$, we find $\bar{\mathcal{F}}_{\mathrm{Bell}}^{(3)}=0.569036$. As before, the pure state biseparability bound extends to mixed states via convexity. Thus, any 3-qubit state for which the combination of (moduli of) expectation values on the right-hand side of Eq.~(\ref{eq:3QB symm average Bell fid}) exceeds $\tfrac{1}{12}\bigl(3+\sqrt{15}\bigr)$ must be genuinely tripartite entangled.

Second, let us turn to the case of $4$ qubits, where we are interested in bounding the quantity
\begin{align}
    \bar{\mathcal{F}}_{\mathrm{Bell}}^{(4)}&=\,\tfrac{1}{24}\Bigl(6
    +|\!\expval{X_{1}X_{2}}\!|+|\!\expval{X_{1}X_{3}}\!|+|\!\expval{X_{1}X_{4}}\!|\nonumber\\[0.5mm]
    &\hspace*{1.05cm}
    +|\!\expval{X_{2}X_{3}}\!|+|\!\expval{X_{2}X_{4}}\!|+|\!\expval{X_{3}X_{4}}\!|\nonumber\\[1mm]
    &\hspace*{1.05cm}
    +\,|\!\expval{Y_{1}Y_{2}}\!|\ +\ |\!\expval{Y_{1}Y_{3}}\!|\,+\,|\!\expval{Y_{1}Y_{4}}\!|\nonumber\\[1mm]
    &\hspace*{1.05cm}
    +\,|\!\expval{Y_{2}Y_{3}}\!|\ +\ |\!\expval{Y_{2}Y_{4}}\!|\,+\,|\!\expval{Y_{3}Y_{4}}\!|\nonumber\\[1mm]
    &\hspace*{1.05cm}
    +|\!\expval{Z_{1}Z_{2}}\!|\,+\,|\!\expval{Z_{1}Z_{3}}\!|\,+\,|\!\expval{Z_{1}Z_{4}}\!|\nonumber\\[0.5mm]
    &\hspace*{1.05cm}
    +|\!\expval{Z_{2}Z_{3}}\!|\,+\,|\!\expval{Z_{2}Z_{4}}\!|\,+\,|\!\expval{Z_{3}Z_{4}}\!|\Bigr).
    \label{eq:4QB symm average Bell fid}
\end{align}
For any pure state that is separable w.r.t. a bipartition into $1$ versus $3$ qubits, we find three triples of expectation values that are ``cut" (each bounded by $1$), while three triples pertaining to the same subsystem can be combined into mutually anticommuting triples, each bounded by $\sqrt{3}$, obtaining
\begin{align}
    \bar{\mathcal{F}}^{1|234}_{\mathrm{Bell}}   &\leq\,\tfrac{1}{24}\bigl(6+3+3\sqrt{3}\bigr)=\tfrac{1}{8}\bigl(3+\sqrt{3}\bigr)\approx0.591506.
    \label{eq:4QB symm average Bell fid bound 1 vs 234}
\end{align}
Instead, we can also use the bound arising from the purity of the reduced state of qubits $234$ of the biseparable pure state. Since the local dimension for these three qubits is $2^{3}=8$ and we have $12$ terms appearing [the Bloch vectors of the $i$th qubit $|\vec{a}_{i}|$ ($i=2,3,4$) and the correlations tensor elements $|t_{nn}^{23}|$, $|t_{nn}^{24}|$, and $|t_{nn}^{34}|$ for $n=1,2,3$], we find the bound
\begin{align}
    \bar{\mathcal{F}}^{1|234}_{\mathrm{Bell}}   &\leq\,\tfrac{1}{24}\bigl(6+12\sqrt{\tfrac{8-1}{12}}\bigr)=\tfrac{1}{8}\bigl(6+\sqrt{84}\bigr)\approx0.631881.
    \label{eq:4QB symm average Bell fid bound 1 vs 234 alternative}
\end{align}
However, this upper bound is larger than that arising just from using the ACT, and Eq.~(\ref{eq:4QB symm average Bell fid bound 1 vs 234 alternative}) is therefore of no further consequence.

The only other possible type of bipartition of 4 qubits is into two sets of 2 qubits. In this case, four triples are cut by the bipartition, but in each set one triple of unpaired operators remains (jointly bounded by $3$), such that we have
\begin{align}
    \bar{\mathcal{F}}^{12|34}_{\mathrm{Bell}}   &\leq\,\tfrac{1}{24}\bigl(6+4+3+3\bigr)=\tfrac{2}{3}.
    \label{eq:4QB symm average Bell fid bound 12 vs 34}
\end{align}
As we have argued before, a biseparability bound for 3 qubits that is larger or equal to $\tfrac{2}{3}$ is not very useful since even pure GME states (e.g., the 4-qubit Dicke state with two excitations) achieve only this value. We hence again turn to using the purity of the subsystems for a biseparable state $\ket{\psi}_{1243}=\ket{\phi}_{12}\ket{\chi}_{34}$. In this case, the symmetric average Bell fidelity can be bounded by
\begin{align}
    \bar{\mathcal{F}}^{12|34}_{\mathrm{Bell}}   &\leq\,\tfrac{1}{24}\bigl(6+
    |\vec{a}_{1}|.|\vec{a}_{3}|+|\vec{a}_{1}|.|\vec{a}_{4}|+|\vec{a}_{2}|.|\vec{a}_{3}|+|\vec{a}_{2}|.|\vec{a}_{4}|
    \nonumber\\[1mm]
    &\ \ \ +\sum\limits_{n=1,2,3}|t_{nn}^{12}|+\sum\limits_{n=1,2,3}|t_{nn}^{34}|\bigr).
    \label{eq:4QB symm average Bell fid bound 12 vs 34 rewritten}
\end{align}
Here, we encounter a different optimisation problem than before, since we no longer seek to maximize the sum of absolute values, but some quantities (e.g., $|\vec{a}_{1}|$ and $|\vec{a}_{3}|$) are coupled. However, due to the symmetric form (w.r.t. the exchange of qubits $12$ with $34$) of the expression, we may write
\begin{align}
    \bar{\mathcal{F}}^{12|34}_{\mathrm{Bell}}   &\leq\,\tfrac{1}{24}\Bigl(6
    +2\bigl(
     |\vec{a}_{1}|^{2}+ |\vec{a}_{2}|^{2}+\!\!\sum\limits_{n=1,2,3}|t_{nn}^{12}|
    \bigr)\Bigr).
    \label{eq:4QB symm average Bell fid bound 12 vs 34 rewritten 2}
\end{align}
We hence seek to maximize $f(a,t)=3t+2a^{2}$ under the constraints $2a^{2}+3t^{2}=3$ and $a^{2}\leq1$, which is achieved for $a=1$ and $t=\tfrac{1}{\sqrt{3}}$, and hence
\begin{align}
    \bar{\mathcal{F}}^{12|34}_{\mathrm{Bell}}   &\leq\,\tfrac{1}{24}\bigl(6
    +2\bigl[2+\tfrac{3}{\sqrt{3}}\bigr]\bigr)=\tfrac{1}{12}\bigl(5+\sqrt{3}\bigr)\approx0.561004.
    \label{eq:4QB symm average Bell fid bound 12 vs 34 rewritten 3}
\end{align}
Since this value is smaller than that for the bipartition $1|234$ in Eq.~(\ref{eq:4QB symm average Bell fid bound 1 vs 234}), we can identify the bound of Eq.~(\ref{eq:4QB symm average Bell fid bound 1 vs 234}) with the biseparability bound for the symmetric average Bell fidelity for $4$ qubits, i.e.,
\begin{align}
    \bar{\mathcal{F}}_{\mathrm{Bell}}^{(4) \mathrm{bisep}}   &\leq
    \tfrac{1}{8}\bigl(3+\sqrt{3}\bigr)\approx0.591506.
    \label{eq:4QB average symm Bell fid bisep bound}
\end{align}

For more than 4 qubits, we can derive more general expressions using the method based on the ACT, while the exponentially increasing subsystem dimension makes bounds based on the subsystem purity unfeasible. Consider a system of $N\geq4$ qubits that is in a separable pure state w.r.t. to a bipartition into a single qubit versus the remaining $N-1$ qubits. One may then identify $N-1$ triples of expectation values that factorize and can be bounded by one, while the remaining $b_{N-1}=b_{N}-(N-1)$ triples form anticommuting sets of three which are each bounded by $\sqrt{3}$. We thus have
\begin{align}
    \bar{\mathcal{F}}^{1|23\ldots N}_{\mathrm{Bell}}   &\leq\tfrac{1}{4}\Bigl(1+\tfrac{1}{b_{N}}\bigl[N-1+\bigl(b_{N}-(N-1)\bigr)\sqrt{3}\bigr]\Bigr)
    \nonumber\\[1mm]
    &\ =\tfrac{1}{4}\bigl(1+\sqrt{3}\bigr)-\tfrac{1}{2N}\bigl(\sqrt{3}-1\bigr),
\end{align}
where we have made use of the fact that $\tfrac{N-1}{b_{N}}=\tfrac{(N-1)2(N-2)!}{N!}=\tfrac{2}{N}$. Note that, as required, this bound reduces to the result obtained in Eq.~(\ref{eq:4QB symm average Bell fid bound 1 vs 234}) for $N=4$. Intuitively, it is now clear that other bipartitions will provide smaller upper bounds, since more operators are affected by the factorization. The exception being the case $N=4$, where we have already seen that the separation into two sets of two provides a larger upper bound since each side then features unpaired expectation values.

To confirm this, let us briefly consider bipartitions into $2$ and $N-2$ qubits for $N\geq5$. In such a case, $2(N-2)$ expectation values factorize for the respective pure, separable states, and one triple of operators pertaining to the two isolated qubits cannot be paired with anticommuting partners, whereas $b_{N-2}=b_{N}-(N-2)=b_{N}-2(N-1)+1$ triples of operators can be matched up in this way. Thus, we have
\begin{align}
    \bar{\mathcal{F}}^{12|34\ldots N}_{\mathrm{Bell}} \! &\leq\tfrac{1}{4}\Bigl(1\!+\!\tfrac{1}{b_{N}}\bigl[
    2(N\!-\!2)\!+\!3\!+\!\bigl(b_{N}\!-\!2N\!+\!3\bigr)\sqrt{3}\,\bigr]\Bigr)
    \nonumber\\[1mm]
    &\ =\tfrac{1}{4}\bigl(1+\sqrt{3}\bigr)\bigl(1+\tfrac{1}{b_{N}}\bigr)
    -\tfrac{1}{N}\bigl(\sqrt{3}-1\bigr).
\end{align}
This expression provides a smaller upper bound when
\begin{subequations}
\label{eq:conditions for 12 vs rest bipartition to be better}
\begin{align}
    \tfrac{1}{2N}\bigl(\sqrt{3}-1\bigr) &>  \tfrac{1}{4b_{N}}\bigl(1+\sqrt{3}\bigr)\\[1mm]
    \Rightarrow\ 1  &>\tfrac{1}{2(N-1)}\bigl(1+\sqrt{3}\bigr)^{2},
\end{align}
\end{subequations}
which is the case for $N\geq5$, as expected. We have also confirmed that this intuition holds for bipartitions into $k$ versus $N-k$ qubits for $3\leq k\leq N-3$. We can hence formulate the biseparability bound based on the symmetric average Bell fidelity for arbitrary numbers of qubits in the following way. For any biseparable state of $N$ qubits, the symmetric average Bell fidelity satisfies
\begin{align}
    \bar{\mathcal{F}}^{(N)}_{\mathrm{Bell}} &\leq\bar{\mathcal{F}}^{(N) \mathrm{bisep}}_{\mathrm{Bell}}:=
    \begin{cases}
        \tfrac{1}{12}\bigl(3+\sqrt{15}\bigr)    &   \mbox{for $N=3$}\\[1mm]
        \tfrac{1}{4}\bigl(1+\sqrt{3}\bigr)-\tfrac{1}{2N}\bigl(\sqrt{3}-1\bigr)  & \mbox{for $N\geq4$}
    \end{cases}.
    \label{eq:N qubit GME witness appendix}
\end{align}
Conversely, any state that violates the inequality Eq.~(\ref{eq:N qubit GME witness appendix}) is genuinely $N$-partite entangled. Before we discuss the practical usefulness of these witnesses, let us briefly analyze possible improvements in Appendix~\hyperref[sec:Optimizing GME Witnesses Based on Bipartite Fidelities]{A.II.4}.


\subsubsection{Optimizing GME Witnesses Based on Bipartite Fidelities}\label{sec:Optimizing GME Witnesses Based on Bipartite Fidelities}

To keep the notation simple during the derivations, we have thus far used only expectation values of pairs of the same Pauli operators, i.e., of the form $|\!\expval{O_{i}O_{j}}\!|$. In practice, this corresponds to measuring the real part of certain off-diagonal elements of the density operator. To see this, consider a $2$-qubit state $\rho$ and note that
\begin{align}
    \tr\bigl(\rho(X_{1}X_{2}+Y_{1}Y_{2})\bigr)  &=\,4\operatorname{Re}(\bra{01}\rho\ket{10}).
    \label{eq:example matrix element}
\end{align}
Of course, the off-diagonal element $\bra{01}\rho\ket{10}$ need not be real for a given $2$-qubit state. Here, one may note that the derivations of all bounds that we have considered so far are invariant under local unitary transformations. That is, we can replace the triple of operators $\{X_{i},Y_{i},Z_{i}\}$ for the $i$th qubit with the rotated operators $\tilde{O}_{i}=U_{i}O_{i}U_{i}^{\dagger}$ for any unitary $U_{i}$. This is the case because such a rotation maps a triple of anticommuting operators to another triple of anticommuting operators, and the length of the Bloch vectors also is left invariant. For instance, one could perform a rotation in the equatorial plane of the Bloch sphere, and map
\begin{align}
    X_{i}  &\mapsto\tilde{X}_{i}\,=\,\cos(\theta_{i})X_{i}\,-\,\sin(\theta_{i})Y_{i}\,,\\[1mm]
    Y_{i}  &\mapsto\tilde{X}_{i}\,=\,\sin(\theta_{i})X_{i}\,+\,\cos(\theta_{i})Y_{i}\,.
    \label{eq:X-Y rotation}
\end{align}
In the example of Eq.~(\ref{eq:example matrix element}) this means we can pick $\theta_{1}=0$ and $\theta_{2}=-\tfrac{\pi}{2}$ to obtain
\begin{align}
    \tr\bigl(\rho(\tilde{X}_{1}\tilde{X}_{2}+\tilde{Y}_{1}\tilde{Y}_{2})\bigr)  &=\,
    \tr\bigl(\rho(X_{1}Y_{2}-Y_{1}X_{2})\bigr)\nonumber\\[1mm]
    &=\,4\operatorname{Im}(\bra{01}\rho\ket{10}).
    \label{eq:example matrix element rotated}
\end{align}
In particular, there exist rotation angles $\theta_{1}$ and $\theta_{2}$ such that
\begin{align}
    \tr\bigl(\rho(\tilde{X}_{1}\tilde{X}_{2}+\tilde{Y}_{1}\tilde{Y}_{2})\bigr)
    &=\,4|\!\bra{01}\rho\ket{10}\!|.
    \label{eq:example matrix element rotated optimal}
\end{align}

In general, one hence has the freedom of $N$ independent transformations $U_{i}\in U(2)$ to optimize the GME witnesses presented so far. In an experimental setting, this optimisation can be done \emph{a priori} if the quantum state $\rho$ that one expects to produce (approximately) in the experiment is known. However, if the underlying state is unknown, one may also measure all combinations of $2$-qubit Pauli operators for all pairs of qubits within the set of $N$ qubits (amounting to $9b_{N}=\tfrac{9}{2}N(N-1)$ $2$-qubit measurements) and perform the optimisation on the experimental data. For instance, the results for $\bar{\mathcal{F}}^{(3)}_{\mathrm{Bell}}$ presented in Fig.~\ref{figure2} of the main text have been obtained by such a postprocessing of available measurement data, and the corresponding optimisation has been restricted to rotations in the $X$-$Y$ planes as shown in Eq.~(\ref{eq:X-Y rotation}).

In addition to \emph{a posteriori} optimisation, one may perform some of these $2$-qubit measurements on different pairs simultaneously if the individual outcomes for each qubit are recorded. For instance, in a register of $N\geq3$ qubits one may obtain the expectation values of $X_{1}X_{2}$ and $X_{2}Y_{3}$ from measuring the first and second qubit in the eigenbasis of $X$ and the third in the eigenbasis of $Y$ and recording all three outcomes in each run. For measurements of this kind, such as have been performed in our experiment, data used to estimate different $2$-qubit expectation values may be correlated, which has to be taken into account in the calculation of the estimate for the variance of the GME witness. This is explained in more detail in Ref.~\cite[Supplementary Information Sections IV.A.4 and IV.A.5 on pages 9--13]{LanyonEtAl2017}.


\subsubsection{Usefulness of GME Witnesses Based on Bipartite Fidelities}\label{sec:Usefulness of GME Witnesses Based on Bipartite Fidelities}

A crucial question when employing these witnesses is of course whether or not states exist that can be detected by them. To analyse this problem, we compare the upper bound $\bar{\mathcal{F}}_{\mathrm{Bell}}^{(N) \mathrm{bisep}}$ for biseparable states with the upper bound $\bar{\mathcal{F}}_{\mathrm{Bell}}^{(N) \mathrm{max}}$ for arbitrary states from Eq.~(\ref{eq:NQB symm average Bell fid bound}) by calculating their distance as a function of the number of qubits. We find the expression
\begin{align}
    \bar{\mathcal{F}}_{\mathrm{Bell}}^{(N) \mathrm{max}}\!-\!\bar{\mathcal{F}}_{\mathrm{Bell}}^{(N) \mathrm{bisep}} \! &=
    \begin{cases}
        \tfrac{1}{12}\sqrt{3}\bigl(3-\sqrt{5}\bigr)    &   \mbox{($N=3$)}\\[1mm]
        \tfrac{1}{2N}\bigl(\sqrt{3}-1\bigr)    &   \mbox{($N\geq4$)}
    \end{cases},
\end{align}
where the numerical values for 3 and 4 qubits are $\tfrac{1}{12}\sqrt{3}\bigl(3-\sqrt{5}\bigr)\approx0.110264$ and $\tfrac{1}{8}\bigl(\sqrt{3}-1\bigr)\approx0.0915064$, respectively. The gap between the bounds is hence largest for $N=3$, and shrinks with increasing number of qubits. It is hence expected that there is some finite $N$ for which no GME states exist that are detected by our witnesses, and at this point, we cannot say for which $N$ this occurs.

For 3 qubits, we have already found examples of genuinely tripartite entangled states that can be detected, i.e., the 3-qubit W state $\ket{\psi_{\mathrm{W}}\suptiny{0}{0}{(3)}}$ and the $2$-excitation Dicke state $\ket{\psi_{\mathrm{D},2}\suptiny{0}{0}{(3)}}$ which provide symmetric average Bell fidelities of $\tfrac{2}{3}$. The experimental results discussed in the main text (see Fig.~\ref{figure2}) further show that $\bar{\mathcal{F}}^{(3)}_{\mathrm{Bell}}$ is also a useful witness for mixed states produced in realistic situations. Beyond 3 qubits, the witnesses $\bar{\mathcal{F}}^{(N\geq4)}_{\mathrm{Bell}}$ (optimized only over rotations in the $X$-$Y$ plane, see Appendix~\hyperref[sec:Optimizing GME Witnesses Based on Bipartite Fidelities]{A.II.4}) have not been able to detect GME in our experimental setting. However, we know that $4$-qubit states exist, e.g., the $4$-qubit $2$-excitation Dicke state $\ket{\psi_{\mathrm{D},2}\suptiny{0}{0}{(4)}}$ which could be detected in this way, since $\bar{\mathcal{F}}^{(4)}_{\mathrm{Bell}}(\ket{\psi_{\mathrm{D},2}\suptiny{0}{0}{(4)}})=\tfrac{2}{3}$. Unfortunately, the $2$-excitation Dicke state for $5$ qubits only provides a value of $\bar{\mathcal{F}}^{(5)}_{\mathrm{Bell}}(\ket{\psi_{\mathrm{D},2}\suptiny{0}{0}{(5)}})=0.6$ whereas $\bar{\mathcal{F}}^{(N) \mathrm{bisep}}_{\mathrm{Bell}}=\tfrac{1}{20}(7+3\sqrt{3})\approx0.61$. Tentative searches for other $5$-qubit states for which $\bar{\mathcal{F}}^{(5)}_{\mathrm{Bell}}$ exceeds the biseparability bound have been unsuccessful thus far. The question of whether GME states exist that can be detected with our method for $N\geq5$ hence remains open.



\renewcommand{\thesubsubsection}{B.\Roman{subsection}.\arabic{subsubsection}}
\renewcommand{\thesubsection}{B.\Roman{subsection}}
\renewcommand{\thesection}{B}
\setcounter{equation}{0}
\numberwithin{equation}{section}
\setcounter{figure}{0}
\renewcommand{\thefigure}{B.\arabic{figure}}

\section{Construction of Witnesses Based on Numerical Search}\label{appendix:ulm witnesses}

\subsection{Genuine Multipartite Negativity}\label{sec:Genuine multipartite negativity}

We now discuss the genuine negativity (GMN) of Ref.~\cite{HofmannMoroderGuehne2014} that we use to quantify GME in the experiment. We present its definition as a convex-roof construction and the alternative way of writing it in terms of a semidefinite program, which turns it into a numerically computable measure of entanglement for an arbitrary mixed state.

We start by introducing the notation and presenting preliminary definitions. Note that a bipartition of $\{ 1,\ldots,N \}$ can be specified by a subset $A \subset \{ 1,\ldots,N \}$ and its complement $\bar{A} = \{ 1,\ldots,N \}\backslash A$. With this, we define the partial transposition on $A$ for an operator $X_A \otimes X_{\bar{A}}$, where $X_A$ and $X_{\bar A}$ act on the Hilbert spaces associated to $A$ and $\bar A$, respectively, as $(X_A \otimes X_{\bar A})^{T_A} = X_A^T \otimes X_{\bar A}$. The definition then extends to any operator on the $N$-particle Hilbert space by linearity. Next, the negativity of a bipartite quantum state $\rho$ with respect to the bipartition $A|\bar A$ is given by the sum of the negative eigenvalues of the partially transposed density matrix, i.e., $\mathcal{N}_{A|\bar A}(\rho) = \sum_{\lambda_i \le 0} |\lambda_i(\rho^{T_A})| = \mathcal{N}_{A|\bar A}(\rho) = \frac{\|\rho^{T_A}\|_1 - 1}{2}$, where $\lambda_i(X)$ denotes the $i$th eigenvalue of an operator $X$. With this, the GMN of an $N$-particle state $\rho$ is given by
\begin{equation}
\mathcal{N}_{\rm g}(\rho) = \min_{\{p_i,\rho_i\}} \sum_{i} p_i \min_{A|\bar A} \mathcal{N}_{A|\bar A}(\rho_i),
\end{equation}
where the inner minimisation is over the possible bipartitions $A|\bar{A}$ of $\{ 1,\ldots,N \}$ and the outer minimisation is over decompositions $\rho = \sum_i p_i \rho_i$, where $\{p_i\}_i$ is a probability distribution and $\rho_i$ are density matrices. For pure states, this reduces to $\mathcal{N}_g(\ket{\psi}\!\bra{\psi}) = \min_{A|\bar A} \mathcal{N}_{A|\bar A}(\ket{\psi}\!\!\bra{\psi})$.

For mixed states the GMN can still efficiently be computed using numerical tools from the field of semidefinite programming. This follows from the fact that the GMN is given by the optimal value of the following optimisation
\begin{equation}
\label{GMN}
\begin{split}
\mathcal{N}_{\mathrm{g}}(\rho)\;\; =\;\; &\max_{Q, P_A, R_A} \; \left(-\tr(Q \rho) \right) \\
& \;\;\;\;\;\;\;\;\; Q = K_A + Q_A^{T_A} \;\;\;\; \forall A|\bar A, \\
& \;\;\;\;\;\;\;\;\; 0 \le K_A \text{ and }0 \le R_A \le \id,
\end{split}
\end{equation}
where $Q, K_A$, and $R_A$ are operators acting on the Hilbert space.
The GMN is zero for all (bi)separable states and, therefore, a nonzero value provides a way to certify GME. More precisely, the GMN is nonzero for any state that cannot be written as a PPT mixture. Recall that a multipartite state $\rho$ is called a PPT mixture if it admits a mixed state decomposition $\rho = \sum_A p_A \rho_A$ where $\{p_A\}_A$ is a probability distribution and $\rho_A$ has a positive partial transposition (we say, $\rho_A$ is PPT) with respect to the bipartition $A|\bar{A}$. That is, formally, $\rho_A^{T_A} \ge 0$. As noted earlier, a state is called biseparable if it can be written as a convex combination of states that are separable with respect to one bipartition $A|\bar A$. Since any separable state is PPT, every biseparable state can be written as a PPT mixture. Consequently, a state with nonzero GMN is GME.

Moreover, the GMN quantifies the entanglement in the sense that a state $\rho$ is more entangled than $\sigma$ if $\mathcal{N}_{\mathrm{g}}(\rho) \ge \mathcal{N}_{\mathrm{g}}(\sigma)$. The underlying mathematical property is that the GMN is nonincreasing under so-called full LOCC operations. In particular, from this property it follows that no GME state can be generated from a non-GME state with local operations only.

A further beneficial property of writing the GMN as in Eq.~\eqref{GMN} is that this yields an entanglement witness, that is, an observable that provides ideally a sharp lower bound to the GMN as in Eq.~\eqref{QEW} in the main text and may be accessed experimentally. For our purposes note that whether a witness can be measured depends on the measurements that are available in our experiment. We therefore discuss those measurements next. The procedure of obtaining the entanglement witnesses that are accessible for us is then described subsequently in Appendix~\hyperref[sec:Accessible QEWs]{B.III} in more detail.

\subsection{Accessible Measurements on $k$ Neighbouring Sites}\label{sec:27 Accessible measurements}

Here, we discuss the operators that can be measured locally on $k$ consecutive sites with the data available in our experiment.
Our starting point is hence the set of all possible observables whose measurement outcomes can be obtained from the 27 measurement settings that we mentioned in Sec.~\ref{sec:initial results}. As we noted there, these measurement settings can be used to obtain estimates of the expectation values of all possible products of the identity and the three Pauli operators on all groups of three neighbouring qubits in the chain of 20 qubits. On such triples of neighbouring qubits this thus allows us to estimate the expectation value of any operator, in particular, any possible entanglement witness. However, the situation is different for more than three neighbouring sites. Regarding this case, recall that in the 27 settings we consider, the 20 qubits are measured simultaneously such that the accessible information turns out to be more than just knowing the three-body reductions of neighbouring qubits. Also as a consequence of how we choose these 27 settings, each of them has the property that the local bases in which sites $i$ and $i+3$ are measured are identical. To illustrate what this implies for the accessible operators, let us consider the example of four sites. In terms of projective qubit measurements, a basis of operators that we can measure with the 27 settings is given by $\mathcal{B}_4 = \{P_{\vec{s},\vec{\alpha}}^{(4)}\}_{\vec{s},\vec{\alpha}}$ where $ P_{\vec{s},\vec{\alpha}}^{(4)}$ denotes the projector onto the state
\begin{equation}
\label{measured basis states}
\ket{\vec{s},\vec{\alpha}} = \ket{s_1,s_2,s_3,s_4}_{\alpha_1,\alpha_2,\alpha_3,\alpha_1},
\end{equation}
where $\vec{s} \equiv (s_1,s_2,s_3,s_4) \in \{\uparrow,\downarrow\}^{\times 4}$ and $\vec{\alpha} \equiv (\alpha_1,\alpha_2,\alpha_3) \in \{1,2,3\}^{\times 3}$. In words, these operators comprise all projective qubit measurements where the first and the fourth qubit are measured in the same direction. Notably, with the Pauli operators $X$, $Y$ and $Z$, this set of operators spans the same subspace as the operators $\{\sigma_{\alpha_1} \otimes \sigma_{\alpha_2} \otimes \sigma_{\alpha_3} \otimes \sigma_{\alpha_1}\}_{\vec{\alpha}}$, where $\vec{\alpha} \in \{0,1,2,3\}^{\times 3}$ and $\sigma_0 = \id$, $\sigma_1 = X$, $\sigma_2 = {Y}$, $\sigma_3 = Z$. For the construction of the witness we will, however, use the projectors as our basis set. Accordingly, generalising this to arbitrary number of sites, we denote the projectors that form a basis of the operators that can be measured on $k$ consecutive sites with the 27 settings by $\mathcal{B}_k$. As a further remark, let us mention that the number of elements in $\mathcal{B}_k$ grows exponentially with $k$ as $|\mathcal{B}_k| = 27 \times 2^k$.

\subsection{Accessible Quantitative Entanglement Witnesses}\label{sec:Accessible QEWs}

Next, we turn to the construction of an entanglement witness that is fully accessible from the available information provided by the 27 measurement settings. The main step is to solve the optimisation given in Eq.~\eqref{accessible bound} below, which provides a quantitative witness. The witness can then be evaluated using the experimental data.

We distinguish between a simple entanglement witness, which is an operator $Q$ that has a positive expectation value $\tr(Q \rho_{\rm sep}) \ge 0$ for any separable state $\rho_{\rm sep}$ and for which there exists at least one entangled state $\rho$ with $\tr(Q \rho) \le 0$, and a \textit{quantitative} entanglement witness, i.e., an operator that fulfills the property of being an entanglement witness and additionally provides a lower bound to the GMN via (minus) its expectation value $\mathcal{S} = -\tr(Q \rho)$ of the form
\begin{equation}
\mathcal{S} \le \mathcal{N}_{\rm g}(\rho)
\end{equation}
for any $\rho$, as noted in the main text.

Now, considering $k$ neighbouring sites in the chain of 20 qubits, the available information from the 27 settings is determined by the projectors in the set $\mathcal{B}_k$ as described above. Then, an entanglement witness $\hat{{W}}$ is accessible from this information if it can be written as a linear combination of operators from $\mathcal{B}_k$, i.e., if
\begin{equation}
\label{QEW projectors}
Q_i^{(k)} = \sum_{\vec{s},\vec{\alpha}} c_{i;\vec{s},\vec{\alpha}}^{(k)} P_{\vec{s},\vec{\alpha}}^{(k)},
\end{equation}
with coefficients $c_{i;\vec{s},\vec{\alpha}}^{(k)} \in \mathbbm{R}$, since, in this case, it is fully determined by the set ${\mathcal{B}}_k$.

Here, for a given state $\rho_i^{(k)}$ (as defined in the main text), we can optimise over the coefficients $c_{i;\vec{s},\vec{\alpha}}^{(k)}$ in order to find a quantitative witness that provides the best lower bound to the GMN of $\rho$. As the computation of the GMN itself, see Eq.~\eqref{GMN}, this optimisation can be expressed as a semidefinite program. That is, with the definition $p_{i,\vec{s},\vec{\alpha}}^{(k)} = \tr(P^{(k)}_{\vec{s},\vec{\alpha}} \rho_i^{(k)})$, the solution of
\begin{equation}
\label{accessible bound}
\begin{split}
\mathcal{S}_i^{(k)}\left(\{ P_{\vec{s},\vec{\alpha}}^{(k)}, p_{i;\vec{s},\vec{\alpha}}^{(k)} \}\right) = &\max_{c_{i;\vec{s},\vec{\alpha}},K_A, R_A} \; \left( - \sum_{\vec{s},\vec{\alpha}} {c}_{i;\vec{s},\vec{\alpha}}^{(k)}{p}_{\vec{s},\vec{\alpha}}^{(k)} \right) \\
& \;\; \sum_{\vec{s},\vec{\alpha}} c_{i;\vec{s},\vec{\alpha}}^{(k)} P_{\vec{s},\vec{\alpha}}^{(k)}  \ge K_A + R_A^{T_A} \;\forall A|\bar A, \\
& \;\; 0 \le K_A \text{ and }0 \le R_A \le \id,
\end{split}
\end{equation}
is the best lower bound of the from $\mathcal{S}_i^{(k)}(\{ P_{\vec{s},\vec{\alpha}}^{(k)}, p_{i;\vec{s},\vec{\alpha}}^{(k)} \}) = -\tr(Q_i^{(k)} \rho_i^{(k)}) \le  \mathcal{N}_{\rm g}(\rho_i^{(k)})$ with $Q_i^{(k)}$ a quantitative witness as in Eq.~\eqref{QEW projectors} and with the coefficients for which the maximum in Eq.~\eqref{accessible bound} is achieved. Note that for brevity we denote both the optimisation parameters in Eq.~\eqref{accessible bound} as well as, in the following, the optimal coefficients by $c_{i;\vec{s},\vec{\alpha}}^{(k)}$.
As a further remark, we note that $\mathcal{S}_i^{(k)}(\{ P_{\vec{s},\vec{\alpha}}^{(k)}, p_{i;\vec{s},\vec{\alpha}}^{(k)} \})$ and the optimal witness $Q$ depend only on the (accessible) probabilities $p_{\vec{s},\vec{\alpha}}$ (and the corresponding projectors), such that these quantities are sufficient to determine the bound.

In order to obtain the bounds in Fig.~\ref{figure3} we determine an entanglement witness $W$ that can be decomposed as a sum of projectors as in Eq.~\eqref{QEW projectors} and, hence, it can readily be evaluated using the frequencies measured in the experiment. The basic steps to determine a quantitative witness are then (as also described in the main text) as follows. We first perform a numerical simulation to obtain the time-evolved state $\rho_{\rm sim}(t)$ and determine its reduced density matrices on neighbouring sites. Then, for every reduction $\rho_{\rm sim}^{\mathcal{X}}(t) = \tr_{\backslash\mathcal{X}}(\rho_{\rm sim}(t))$, where $\mathcal{X}\subset \{1,\ldots,N\}$ denotes the sites corresponding to the Hilbert space on which $\rho_{\rm sim}^{\mathcal{X}}(t)$ acts, we solve the optimisation of Eq.~\eqref{accessible bound} (plus some practical amendments, see Appendix~\hyperref[sec:Further constraints]{B.IV}) with $p_{i;\vec{s},\vec{\alpha}}^{(k)} = \tr(P_{\vec{s},\vec{\alpha}}^{(k)} \rho_{\rm sim}^{\mathcal{X}}(t) )$ as input to obtain the witness.

In the numerical simulation we include mixing due to an imperfect initial state, as described in the main text. To this end, we use an initial state that is diagonal in the $Z$ bases as $\rho_{\rm sim}(0) = \sum_{\vec{s}} f_{\vec{s}} \ket{\vec{s}}\!\!\bra{\vec{s}}$, where $\ket{\vec{s}} \equiv \ket{\vec{s},(3,\cdots,3)}$ is a state of the product basis in $Z$ direction (see Eq.~\eqref{measured basis states} for the notation we use) and $f_{\vec{s}}$ is the frequency of the qubit configuration $\vec{s}$ that we observe using the experimental data obtained at $t = 0$~ms. Since, besides the ideal initial state with alternating qubits, only 42 other configurations with nonzero frequency $f_{\vec{s}}$ occur, the participating configurations are readily evolved separately and mixed in order to obtain a density matrix $\rho_{\rm sim}(t)$ at later times.

The optimisation of Eq.~\eqref{accessible bound} used to determine the QEW $Q^{(k)}_i$ depends on the probabilities $p_{i;\vec{s},\vec{\alpha}}^{(k)}$ but not on the underlying quantum state $\rho^{(k)}_i$.
One may be tempted to insert experimentally measured estimates for the probabilities $p_{i;\vec{s},\vec{\alpha}}^{(k)}$ to determine a witness which is optimal for the (unknown) quantum state of the experiment.
However, there is, in general, no quantum state $\rho^{(k)}_i$ which satisfies $\tr(P_{\vec{s},\vec{\alpha}}^{(k)} \rho^{(k)}_i) = \tilde p_{i;\vec{s},\vec{\alpha}}^{(k)}$ where the $\tilde p_{i;\vec{s},\vec{\alpha}}^{(k)}$ are experimentally estimated probabilities because the $\tilde p_{i;\vec{s},\vec{\alpha}}^{(k)}$ are affected by statistical noise from a finite number of measurements.
As a consequence, the optimisation may fail to be feasible.
Therefore, we use the probabilities $p_{i;\vec{s},\vec{\alpha}}^{(k)}$ from the reduced density matrices of the numerically simulated state $\rho_\text{sim}(t)$ in the optimisation of Eq.~\eqref{accessible bound} in order to determine the QEW $Q^{(k)}_i$.
After $Q^{(k)}_i$ has been determined, its expectation value, which provides the lower bounds in Fig.~\ref{figure3}, is determined from the experimental data.

Instead of using the numerically simulated state $\rho_{\rm sim}(t)$ to determine the witness, we furthermore test states obtained from quantum state tomography. For this purpose we split the data into two statistically independent sets of samples and use one set for tomography and the other to evaluate the witness. We pursue this strategy using global pure state MPS reconstruction and find no significant advantage of the approach over using the numerically simulated state $\rho_\text{sim}(t)$ as a starting point.

After performing the optimisation of Eq.~\eqref{accessible bound} to determine $Q^{(k)}_i$, we perform an additional optimisation which makes the QEW more robust to statistical noise and experimental imperfection.
The lower bounds presented in Fig.~\ref{figure3} of the main text are based on this improved witness and the remaining steps are discussed in the next section.

\subsection{Additional Conditions for Practical Witnesses}\label{sec:Further constraints}

To make the witnesses less sensitive to experimental imperfections, we modify the coefficients $c_{i;\vec{s},\vec{\alpha}}^{(k)}$ as discussed in the following.
The major step is to reduce the number of nonzero coefficients. This can be achieved by employing an interactive $l_1$-relaxation method with weight updating that is intended to find sparse solutions \cite{CandesWakinBoyd2008}. We incorporate this into our considerations by solving the following optimisation in the $n$th iteration step,
\begin{equation}
\label{l1 minimization}
\begin{split}
&\min_{c_{i;\vec{s},\vec{\alpha}}^{(k),n}} \; \sum_{i,\vec{s},\vec{\alpha}} \frac{|c_{i;\vec{s},\vec{\alpha}}^{(k),n}|}{|c_{i;\vec{s},\vec{\alpha}}^{(k),n-1}|+\epsilon} \\
& \;\;\;\;\;\;\;\;  \tr\left[\left({Q}^{(k),n}_i -  {Q}^{(k),0}_i\right)\rho_i^{(k)}\right] \le \epsilon, \\
& \;\;\;\;\;\;\;\; {Q}^{(k),n}_i \ge K_A + R_A^{T_A}  \;\;\;\; \forall A|\bar A,
\end{split}
\end{equation}
where $Q^{(k),n}_i = \sum_{\vec{s},\vec{\alpha}} c_{i;\vec{s},\vec{\alpha}}^{(k),n} {P}_{\vec{s},\vec{\alpha}}^{(k)}$, and initially the coefficients are obtained from Eq.~\eqref{accessible bound} with $c_{\vec{s},\vec{\alpha}}^{(k),0} = c_{i;\vec{s},\vec{\alpha}}^{(k)}$ with the coefficients that are found from solving Eq.~\eqref{accessible bound}. We allow the bound to diminish by at most $\epsilon$ from the optimal one in order to find a sparse solution. In the results we present, we choose $\epsilon = 5\times 10^{-3}$ and use three iteration steps. Except for the tolerance $\epsilon$, this step does not deteriorate the bound.
We then further add two simple constraints to Eqs.~\eqref{accessible bound} and~\eqref{l1 minimization}. First, we choose the coefficients as $c_{i;\vec{s},\vec{\alpha}}^{(k),n} \in [-1,1]$. Second, we add the semidefinite condition $K_A \le \id$ to Eq.~\eqref{accessible bound}. Note that this condition has previously been employed in an earlier version of the GMN, see Refs.~\cite{JungnitschMoroderGuehne2011, HofmannMoroderGuehne2014}. However, there the quantitative witnesses are exact functions of the operators $K_A$ and $R_A$, whereas here they are only related via an inequality.
Solving Eqs.\ \eqref{accessible bound} and \eqref{l1 minimization} with the constraints introduced in this section results in the bounds $\mathcal{S}_{i}^{(k)}$ presented in the main text. We observe that the additional modifications significantly help to improve the lower bounds.

\end{document}